\documentclass[twocolumn,epjc3]{svjour3}
\usepackage{graphics}
\usepackage{microtype}
\usepackage{bbm}
\usepackage{amsmath}
\usepackage{mathtools}
\usepackage{caption}
\usepackage{subcaption}
\usepackage{amssymb}
\usepackage{mathrsfs}       
\usepackage[pdftex]{graphicx}
\usepackage{pgfplots}
\pgfplotsset{width=10cm,compat=1.9}
\usepackage{xargs}       
\usepackage{wasysym}
\usepackage{engrec}
\usepackage{enumerate}
\usepackage{appendix}
\usepackage{empheq}
\usepackage{float}
\usepackage{stmaryrd}
\usepackage[parfill]{parskip}
\usepackage[pdftex,breaklinks]{hyperref}
\usepackage{titletoc}
\usepackage{makecell}
\usepackage{lscape}
\usepackage{algorithm}
\usepackage{algorithmicx}
\usepackage{tensor}
\usepackage{algpseudocode}
\usepackage{blkarray}
\usepackage{multirow}
\usepackage{bigstrut}
\usetikzlibrary{plotmarks}

\newcommand{\inv}[1]{\frac 1{#1}}
\newcommand{\ket}[1]{|#1\rangle}
\newcommand{\bra}[1]{\langle#1|}
\newcommand{\braket}[2]{\langle#1|#2\rangle}

\newcommand{\nucl}[2]{\({}^{#2}\mathrm{#1}\)}

\newenvironment{customlegend}[1][]{%
    \begingroup
    \csname pgfplots@init@cleared@structures\endcsname
    \pgfplotsset{#1}%
}{%
    \csname pgfplots@createlegend\endcsname
    \endgroup
}%
\def\addlegendimage{\csname pgfplots@addlegendimage\endcsname}

\begin{document}
\title{Multi-reference many-body perturbation theory for nuclei}
\subtitle{III. Ab initio calculations at second order in PGCM-PT}
\author{M. Frosini\thanksref{ad:saclay,em:mf} \and T. Duguet\thanksref{ad:saclay,ad:kul,em:td} \and J.-P. Ebran\thanksref{ad:dam,ad:fakedam,em:jpe} \and B.~Bally\thanksref{ad:dft,em:bb} \and H.~Hergert\thanksref{ad:msu1,ad:msu2,em:hh} 
\and T.R.~Rodr\'iguez\thanksref{ad:dft,ad:cdiaf,em:tr} \and R.~Roth\thanksref{ad:tud,ad:hfhf,em:rr} \and J. M. Yao\thanksref{ad:sysu,em:jmy}  \and V. Som\`a\thanksref{ad:saclay,em:vs}}

\institute{
\label{ad:saclay}
IRFU, CEA, Universit\'e Paris-Saclay, 91191 Gif-sur-Yvette, France 
\and
\label{ad:kul}
KU Leuven, Department of Physics and Astronomy, Instituut voor Kern- en Stralingsfysica, 3001 Leuven, Belgium 
\and
\label{ad:dam}
CEA, DAM, DIF, 91297 Arpajon, France
\and
\label{ad:fakedam}
Universit\'e Paris-Saclay, CEA, Laboratoire Mati\`ere en Conditions Extr\^emes, 91680, Bruy\`eres-le-Ch\^atel, France
\and
\label{ad:dft}
Departamento de F\'isica Te\'orica, Universidad Aut\'onoma de Madrid, 28049 Madrid, Spain
\and
\label{ad:msu1}
Facility for Rare Isotope Beams, Michigan State University, East Lansing, MI 48824-1321, USA
\and
\label{ad:msu2}
Department of Physics and Astronomy, Michigan State University, East Lansing, MI 48824-1321, USA
\and
\label{ad:cdiaf}
Centro de Investigaci\'on Avanzada en F\'isica Fundamental-CIAFF-UAM, 28049 Madrid, Spain
\and
\label{ad:tud}
Institut f\"ur Kernphysik, Technische Universit\"at Darmstadt, 64289 Darmstadt, Germany
\and
\label{ad:hfhf}
Helmholtz Forschungsakademie Hessen f\"ur FAIR, GSI Helmholtzzentrum, 64289 Darmstadt, Germany
\and  
\label{ad:sysu}
School of Physics and Astronomy, Sun Yat-sen University, Zhuhai 519082, P.R. China
}

\thankstext{em:mf}{\email{mikael.frosini@cea.fr}}
\thankstext{em:td}{\email{thomas.duguet@cea.fr}}
\thankstext{em:jpe}{\email{jean-paul.ebran@cea.fr}}
\thankstext{em:bb}{\email{benjamin.bally@uam.es}}
\thankstext{em:hh}{\email{hergert@frib.msu.edu}}
\thankstext{em:tr}{\email{tomas.rodriguez@uam.es}}
\thankstext{em:rr}{\email{robert.roth@physik.tu-darmstadt.de}}
\thankstext{em:jmy}{\email{yaojm8@mail.sysu.edu.cn }}
\thankstext{em:vs}{\email{vittorio.soma@cea.fr}}

\date{Received: \today{} / Revised version: date}

\maketitle
%
%
\begin{abstract}
In spite of missing dynamical correlations, the projected generator coordinate method (PGCM) was recently shown to be a suitable method to tackle the low-lying spectroscopy of complex nuclei. Still, describing absolute binding energies and reaching high accuracy eventually requires the inclusion of dynamical correlations on top of the PGCM. In this context, the present work discusses the first realistic results of a novel multi-reference perturbation theory (PGCM-PT) that can do so within a symmetry-conserving scheme for both ground and low-lying excited states. First, proof-of-principle calculations in a small ($e_{\mathrm{max}}=4$) model space demonstrate that exact binding energies of closed- (\nucl{O}{16}) and open-shell (\nucl{O}{18}, \nucl{Ne}{20}) nuclei are reproduced within $0.5-1.5\%$ at second order, i.e. through PGCM-PT(2). Moreover, profiting from the pre-processing of the Hamiltonian via multi-reference in-medium similarity renormalization group transformations, PGCM-PT(2) can reach converged values within smaller model spaces than with an unevolved Hamiltonian. Doing so, dynamical correlations captured by PGCM-PT(2) are shown to bring essential corrections to low-lying excitation energies that become too dilated at leading order, i.e., at the strict PGCM level. The present work is laying the foundations for a better understanding of the optimal way to grasp static and dynamical correlations in a consistent fashion, with the aim of accurately describing ground and excited states of complex nuclei via ab initio many-body methods.
\end{abstract}

\section{Introduction}
\label{intro}

The recent breaking of ab initio calculations away from p-shell nuclei into the realm of mid-mass nuclei has been made possible by the formulation and implementation of so-called {\it many-body expansion methods}. Because of their polynomial scaling with system size, expansion methods provide the best candidates yet to extend the reach of ab initio calculations to even heavier nuclei. However, and as explained in the introduction to the first paper of the present series~\cite{frosiniI}, hereafter referred to as Paper I, a current challenge concerns the optimal way to consistently capture both static and dynamical correlations within such methods. While doubly closed-shell nuclei are dominated by (weak) dynamical correlations that are efficiently grasped through a coherent sum of (mostly low-rank) particle-hole excitations of a symmetry-conserving unperturbed product state, open-shell nuclei display strong static correlations that cannot be conveniently accounted for in this way. This results in the necessity to design expansion methods based on more general unperturbed states that can already capture static correlations.

A good candidate to provide appropriate unperturbed states is the projected generator coordinate method (PGCM). The main conclusion of the second paper of the present series~\cite{frosiniII} (Paper II in the following) is that the PGCM is suitable to address the low-lying spectroscopy of complex nuclei within reasonable theoretical uncertainties in spite of missing dynamical correlations. For instance, the energy spectrum and electric multipole transition strengths of the low-lying parity-doublet bands in $^{20}$Ne were reproduced by taking into account both quadrupole and octupole collective fluctuations.

Still, describing absolute binding energies, accounting consistently for a wide range of spectroscopic observables, tackling a large class of nuclei displaying different characteristics and achieving high accuracy eventually requires the inclusion of dynamical correlations on top of the PGCM. This coherent incorporation is made possible by expanding the {\it wave operator} $\Omega$ connecting the PGCM state to the exact eigenstate via the novel multi-reference perturbation theory (PGCM-PT) formulated in Paper I. Doing so, PGCM-PT embeds, for the first time, the PGCM within a systematic {\it symmetry-conserving} expansion method.  

The objective of the present work, the third paper of the series, is to discuss first proof-of-principle results of second-order PGCM-PT, i.e. PGCM-PT(2), calculations in three selected nuclei, namely the doubly closed-shell \nucl{O}{16}, the singly open-shell \nucl{O}{18} and the doubly open-shell \nucl{Ne}{20} that was studied at length at the PGCM level in Paper II.

In addition to displaying the first set of PGCM-PT(2) results, the goal of the present work is to do so while exploiting an additional degree of freedom at our disposal in quantum many-body calculations, i.e. the possible pre-processing of the Hamiltonian, e.g., via unitary transformations generated by nucleus-dependent in-medium similarity renormalization group (IMSRG) evolution. While nucleus-independent vacuum similarity renormalization group (VSRG) transformations of the Hamiltonian have already become a standard tool to pre-sum ultra-violet (UV) dynamical correlations by decoupling  low- and high-momentum modes, nucleus-dependent transformations can be exploited more systematically to pre-sum infra-red (IR) dynamical correlations. 

\begin{figure*}
    \centering
    \includegraphics[width=\textwidth]{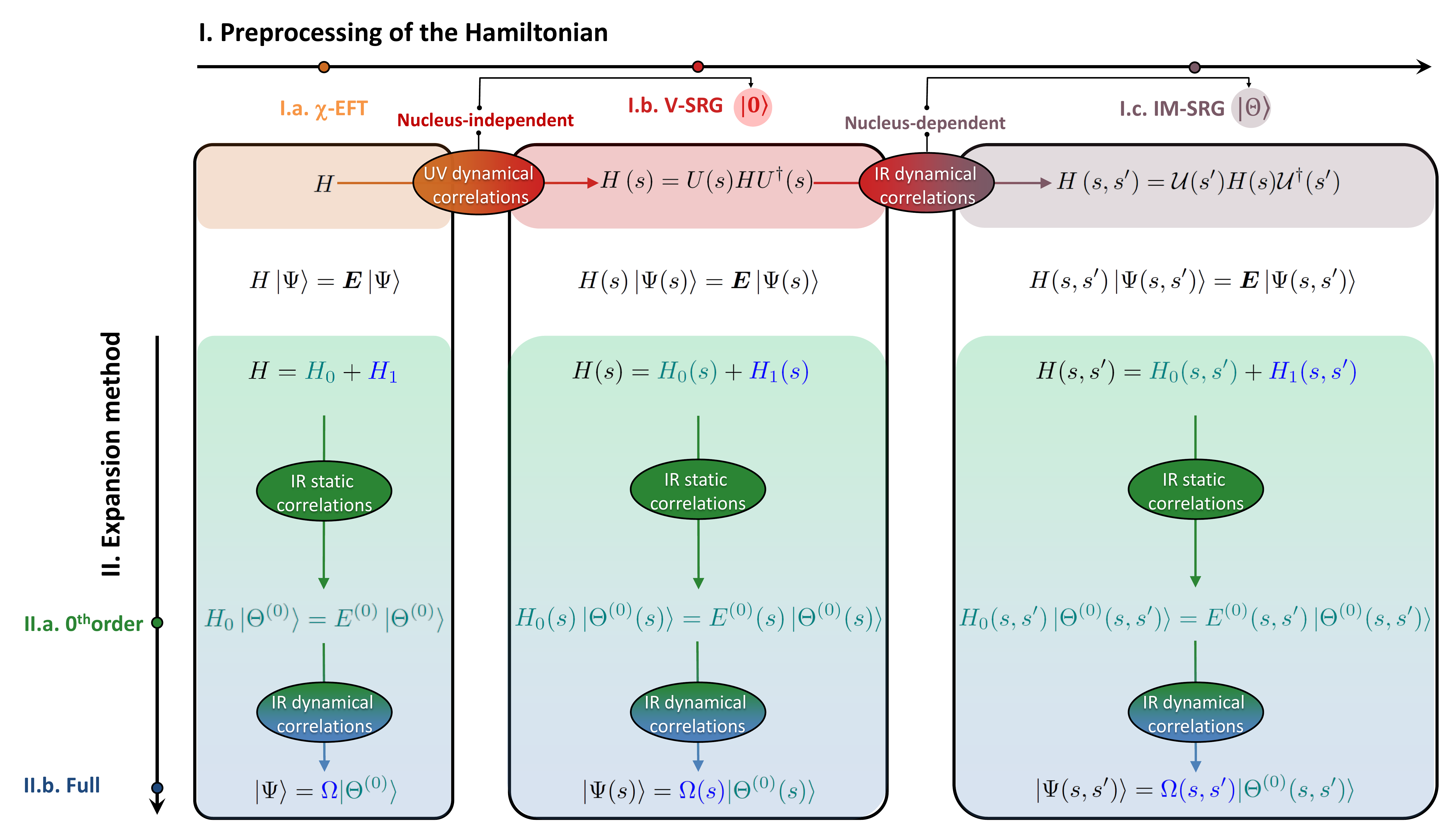}
    \caption{(color online) Schematic workflow of expansion many-body methods (vertical axis) versus potential pre-processings of the Hamiltonian (horizontal axis). Unitary vacuum (in-medium) similarity renormalization group transformations denote a nucleus-independent (nucleus-dependent)  pre-processing of the Hamiltonian.}
    \label{fig:workflow}
\end{figure*}

The single-reference IMSRG (SR-IMSRG) method~\cite{Tsukiyama:2010rj,Hergert:2015awm} applicable to closed-shell systems can in principle fully decouple the unperturbed product state from the rest of the Hilbert space, i.e. from the ${\cal Q}$ space, and thereby make it the actual ground-state of the pre-processed Hamiltonian at the end of the flow. In this case, the wave operator eventually becomes nothing but the identity operator and the expansion method on top of the unperturbed state is trivial. The more general multi-reference IMSRG (MR-IMSRG) method~\cite{Hergert:2014iaa,Hergert:2016iju,Hergert:2016etg} applicable to all nuclei cannot, even in principle, fully decouple the PGCM unperturbed state from the associated ${\cal Q}$ space such that non-zero dynamical correlations remain to be included via a non-trivial, e.g. PGCM-PT, wave operator\footnote{While the IMSRG constitutes per se a method to solve Schr{\"o}dinger's equation when the SR-IMSRG implementation can be applied, it is not the case for the MR-IMSRG approach that can only be seen as a pre-processing of the Hamiltonian on top of which an appropriate many-body method must be applied.}. While the impact of these remaining dynamical correlations on absolute energies may be small, we will show that their inclusion is important for a proper description of low-lying excitation spectra. 

Eventually, a clear picture will emerge that is schematically illustrated in Fig.~\ref{fig:workflow}. Three complementary levers must be consistently exploited to efficiently capture correlations within many-body expansion methods in order to describe (complex) nuclei
\begin{enumerate}
\item the pre-processing of the Hamiltonian,
\item the possibly non-trivial nature of the unperturbed state,
\item the rationale of the expansion.
\end{enumerate}
While each lever is best suited to capture a certain category of correlations, the latter are not orthogonal to one another such that the ideal way to share the load is unclear and will require extensive trial-and-error in the future. The present work wishes to contribute to this long-term endeavor. 

The present paper is organized as follows. Section~\ref{VSRG_H} details the results obtained in \nucl{O}{16}, \nucl{O}{18} and  \nucl{Ne}{20} on the basis of (the two-body part of) a chiral effective field theory ($\chi$EFT) Hamiltonian evolved via VSRG. Section~\ref{IMSRG_H} then elaborates on the impact of further pre-processing the Hamiltonian via MR-IMSRG transformations on the results. Conclusions and future perspectives are eventually discussed in Sec.~\ref{conclusions}. A set of technical appendices provides additional details about the numerical solution of the large-scale linear system of equations at play in PGCM-PT(2) calculations. 

\section{Calculations with VSRG pre-processing}
\label{sec:1}

The reader is referred to Paper I for all necessary details about the PGCM-PT formalism as well as to Refs.~\cite{Bogner:2009bt,PhysRevLett.107.072501,PhysRevC.90.024325} and Refs.~\cite{Hergert:2015awm,Hergert:2016iju,Hergert:2016etg} for vacuum and in-medium IMSRG methods, respectively. The PGCM-PT(2) solver is built on top of an axially-deformed Hartree Fock Bogoliubov (HFB) code~\cite{frosini21e} and a consistent PGCM solver~\cite{frosini21f} allowing for the projections on good particle number, angular momentum and parity. All notations used below are consistent with those introduced in Papers I and II that should be consulted for reference. 

\subsection{Numerical setting}
\label{VSRG_H}

Proof-of-principle calculations are performed using the spherical harmonic oscillator (HO) basis of the one-body Hilbert space ${\cal H}_1$ characterized by an oscillator frequency  \(\hbar\omega=20\text{ MeV} \) and \(5\) oscillator shells ($e_{\text{max}} = 4$). 

The next-to-next-to-next-to-leading order (N$^3$LO) $\chi$EFT Hamiltonian introduced in Refs.~\cite{Huther_2020,Entem:2017gor} and evolved via VSRG to the low-momentum resolution scale \(\lambda_{\text{vsrg}}=1.88\text{ fm}^{-1}\) is employed. Thus, UV dynamical correlations are already processed via the VSRG decoupling of low- and high-momentum modes. 

In these proof-of-principle calculations, only the two-body part of the evolved Hamiltonian is actually retained. Thus, the goal is not to reproduce experimental data but rather to benchmark PGCM-PT(2) results against those obtained from full configuration interaction (FCI) calculations in the same $e_{\text{max}} = 4$ space. The FCI calculations rely on a sequence of $N_{\mathrm{max}}$-truncated many-body Hilbert spaces up to \(N_{\mathrm{max}}=8\) embedded into the FCI space defined by the $e_{\text{max}} = 4$ truncation on the single-particle basis. The results are extrapolated to the full $e_{\text{max}} = 4$ model space limit such that FCI results come with an uncertainty associated with this extrapolation\footnote{The uncertainties on excitation energies do not originate from this extrapolation but are taken from the difference between the results obtained for the largest \(N_{\mathrm{max}}=8\) and the smallest space. Excitation energies are more accurate than absolute ones because they converge faster with $N_{\mathrm{max}}$.}. 

Additional many-body methods are also considered for comparisons. First, the sub-cases of PGCM and PGCM-PT obtained by only using one "seed" HFB state, i.e. omitting the GCM part of the calculation, are considered and referred to as PHFB and PHFB-PT methods. The case where the projection part is further omitted is utilized as well. This single-reference limit of PGCM-PT has been formally elaborated on in Paper I and denotes a symmetry-breaking scheme in case the seed state (i.e. becoming the unperturbed state) is itself symmetry breaking. This limit will be also compared to the standard single-reference symmetry-breaking Bogoliubov many-body perturbation theory (BMBPT)~\cite{Duguet:2015yle,Tichai18BMBPT,Arthuis:2018yoo,Demol:2020mzd,Tichai2020review}.

Our study focuses on three nuclei of increasing complexity. In each case, a different collective coordinate $q$ is employed at the constrained HFB (cHFB) level and for the subsequent GCM mixing\footnote{In each case, the employed interval of $q$ values ensures the convergence of the PGCM calculation with respect to that degree of freedom.}. The characteristics of the associated mean-field, PGCM and PGCM-PT calculations are
\begin{enumerate}
\item Doubly closed-shell \nucl{O}{16}
\begin{itemize}
\item spherically-symmetric Hartree-Fock (HF) states,
\item constraint on the root-mean-square (rms) matter radius ($q\equiv r_{\text{rms}}$),
\item no symmetry projection needed.
\end{itemize}
\item Singly open-shell \nucl{O}{18}
\begin{itemize}
\item spherically-symmetric HFB states,
\item constraint on the pairing gap  ($q\equiv \delta$)~\cite{Duguet:2020hdm},
\item projection of neutron number $N$.
\end{itemize}
\item Doubly open-shell \nucl{Ne}{20}
\begin{itemize}
\item axially-deformed HFB states,
\item constraint on the axial quadrupole moment  ($q\equiv q_{20}$),
\item projections on neutron $N$ and proton $Z$ numbers as well as on angular momentum $J$.
\end{itemize}
\end{enumerate}

\subsubsection{\nucl{O}{16}}

\begin{figure}
    \centering
    \includegraphics[width=.5\textwidth]{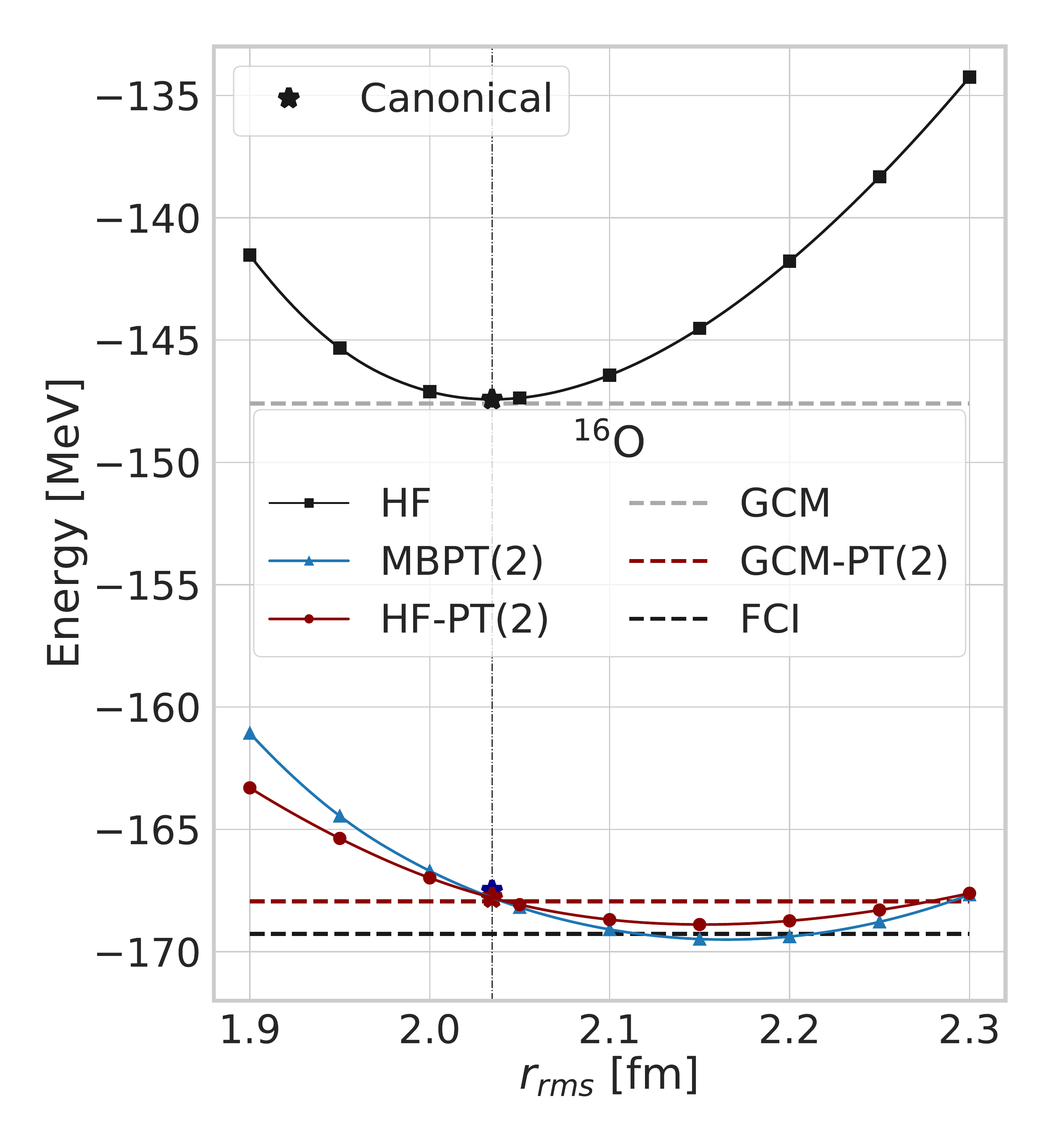}
    \caption{(Color online) Ground-state energy of \nucl{O}{16}  as a function of $r_{\text{rms}}$ of the (underlying) HF vacua.}
    \label{fig:O16_pgcmt}
\end{figure}

In doubly closed-shell systems, the mean-field solution is nothing but a spherical HF state. Since the rms radius operator employed to perform constrained calculations commutes with the total angular momentum $J^2$, all mean-field states involved in the \nucl{O}{16} calculation carry good symmetry quantum numbers and no symmetry projection is necessary in the subsequent PGCM and PGCM-PT calculations, hence they are simply referred to as GCM and GCM-PT, respectively. 

The ground-state total energy curves (TECs) of \nucl{O}{16} are displayed in Fig.~\ref{fig:O16_pgcmt} as a function of the rms radius $r_{\text{rms}}$ of the (underlying) HF vacua. One first observes that cHF and GCM results are underbound by about 20\,MeV ($12\%$) with respect to FCI, missing significant IR dynamical correlations. In the present case\footnote{The GCM and GCM-PT(2) calculations are performed on the basis of the nine cHF states visible on the TEC.}, the GCM adds almost no energy (specifically, $165$\,keV) to the HF minimum, which signals that static IR collective correlations are marginal in such a doubly closed-shell nucleus.

Given the negligible character of static correlations, \nucl{O}{16} acts as a good benchmark for the (P)GCM-PT formalism. First, its single-reference reduction HF-PT(2) is, as formally demonstrated in Paper I, identical to {\it canonical} MBPT(2), i.e. M{\o}ller-Plesset MBPT based on the unconstrained HF solution at the minimum of the TEC  ($r_{\text{rms}}=2.03$\,fm). While both single-reference partitionings of the Hamiltonian provide slightly different results away from the minimum of the HF TEC, they are qualitatively and quantitatively similar. The minima of the two TECs are close to the FCI result. However, perturbation theories are not variational such that it is difficult to argue that these values are to be preferred to canonical ones. As a matter of fact MBPT(3) (not shown) does not flatten the curve in the vicinity of the lowest MBPT(2) value\footnote{Empirically, the MBPT expansion shows less sign of convergence away from the canonical point such that the corresponding MBPT(2) values should not be preferred.}. 

\begin{figure}
    \centering
    \includegraphics[width=.5\textwidth]{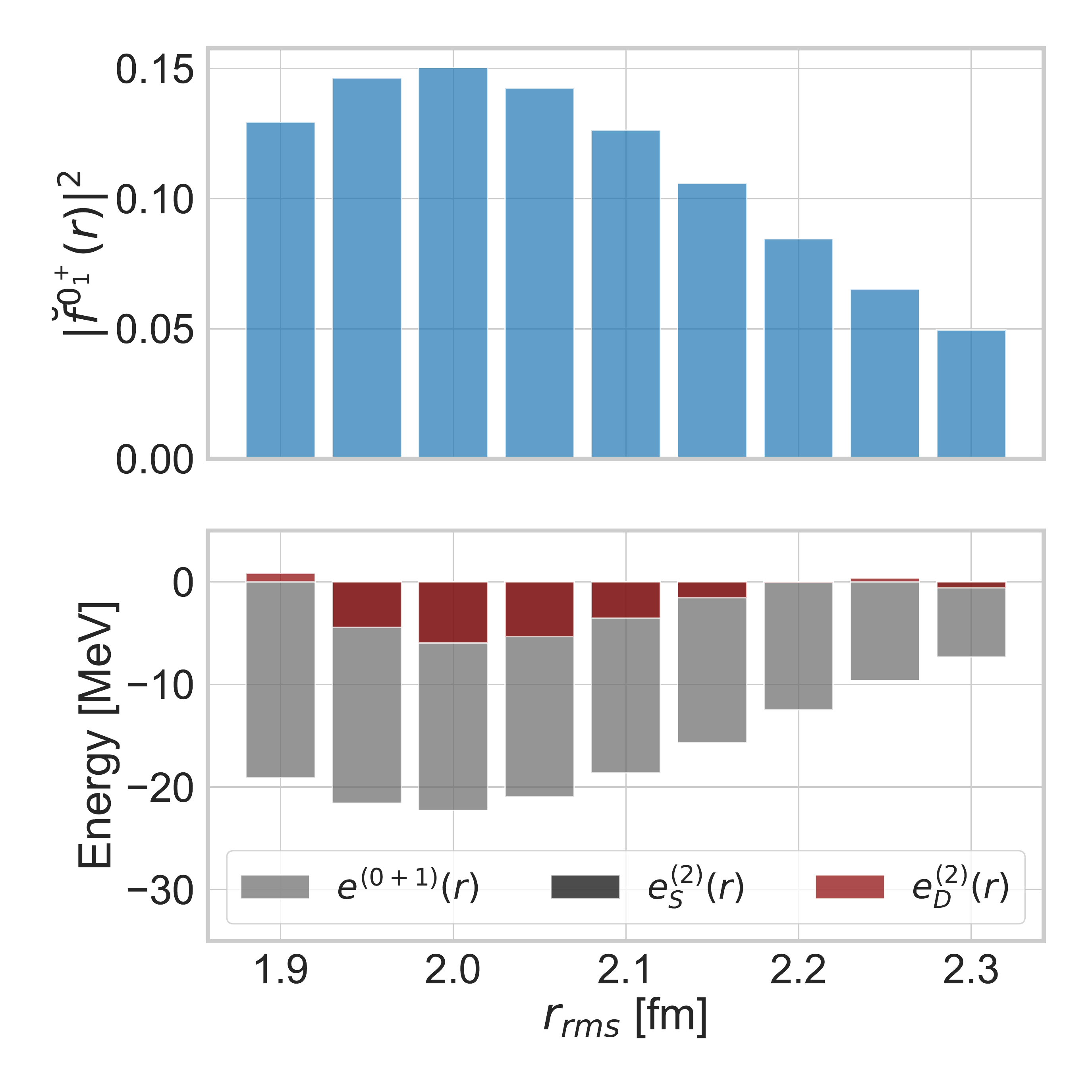}
    \caption{(Color online) Upper panel: collective ground-state GCM wave-function probability distribution ($|\breve{f}^{0^+_1}(r_{\text{rms}})|^2$) in \nucl{O}{16} as a function of the rms radius of the underlying HF vacua. Lower panel: contributions to GCM ($e^{(0+1)}_{0}(r_{\text{rms}})$) and GCM-PT(2) ($e^{(2)}_{S}(r_{\text{rms}})+e^{(2)}_{D}(r_{\text{rms}})$) ground-state energies as a function of $r_{\text{rms}}$. GCM-PT(2) constributions are split into single (one-particle/one-hole) and double (two-particle/two-hole) excitations.}
    \label{fig:O16_wave_functions}
\end{figure}

Focusing on the canonical point, one observes that GCM-PT(2) is consistent with MBPT(2)/HF-PT(2), adding only $146$\,keV static correlation energy. This consistency constitutes a validation of GCM-PT(2), knowing that it is formally very different from MBPT(2) and relies on a completely different numerical procedure as can be appreciated from the various appendices to the present paper. 

Furthermore, this consistency sheds some light on single-reference MBPT(2)/HF-PT(2) results. The upper panel of Fig.~\ref{fig:O16_wave_functions} shows that, while the GCM ground-state collective wave-function spreads over a large interval of $r_{\text{rms}}$ values due to nuclear-size fluctuations, the Hamiltonian dictates that the contributions to the left of the HF minimum (i.e. for $r_{\text{rms}}\le 2.03$\,fm) dominate it. From the energetic viewpoint, the lower panel  of Fig.~\ref{fig:O16_wave_functions}, which shows the decomposition\footnote{The PGCM collective wave function and the contribution $e^{(0+1)}_{0}(r_{\text{rms}})$ of each value of the collective coordinate to the PGCM energy are introduced in App. A of Paper II.} of the GCM energy  as a function of $r_{\text{rms}}$  demonstrates that the largest contributions originate from configurations centered around the HF minimum. Next\footnote{The decomposition of the PGCM-PT(2) correlation energy is provided in Sec. 3.3.2 of Paper I.}, the lower panel also illustrates that the physically-informed weights in the GCM unperturbed state propagate to GCM-PT(2) such that configurations around the HF minimum contribute the most to the second-order correction whereas those associated with the lowest MBPT(2)/HF-PT(2) values around $r_{\text{rms}}\in [2.1,2.2]$\,fm are largely subleading. Eventually, the total GCM-PT(2) energy is nearly identical to canonical MBPT(2)/HF-PT(2) results. This definitely gives more credit to low-order MBPT(2)/HF-PT(2) energies obtained at the canonical point than to those obtained at smaller and larger values of $r_{\text{rms}}$. Interestingly, one also observes that the GCM-PT(2) energy correction is dominated by double (two-particle/two-hole) excitations given that the energy contribution of single (one-particle/one-hole) excitations is negligible at all values of  $r_{\text{rms}}$. While this feature is expected at the canonical point given that single excitations do not contribute to MBPT(2)/HF-PT(2)\footnote{This a consequence of Brillouin's theorem that implies a decoupling between the HF reference state and its singles excitations at the canonical point.}, it is not evident away from it.

Eventually, the GCM-PT(2) binding energy differs by $0.8\%$ from the FCI result. A common theme throughout the paper regards the best way to achieve even greater accuracy. At this point, one can either hope to enrich the PGCM unperturbed state by selecting a potentially pertinent additional collective coordinate $q$ and/or go to PGCM-PT(3)\footnote{The benefit of going to PGCM-PT(3) (see Ref.~\cite{Ladoczki20} for a similar situation) given the associated numerical scaling makes probably more efficient to seek for a further improvement of the PGCM unperturbed state.}. A third (complementary) option to achieve such a goal will be introduced in Sec.~\ref{IMSRG_H}.

\subsubsection{\nucl{O}{18}}
\label{O182body}

\begin{figure}
    \centering
    \includegraphics[width=.5\textwidth]{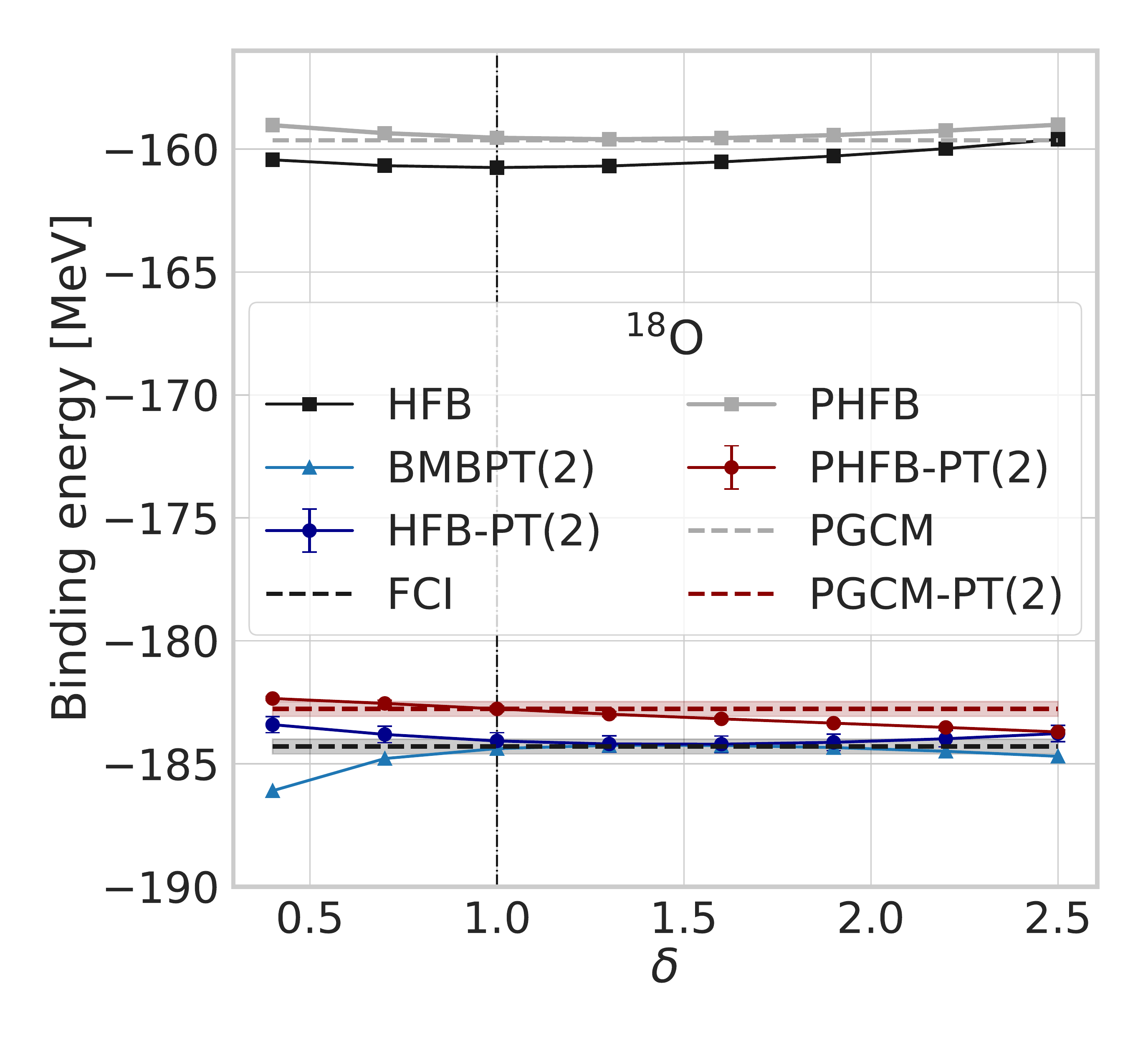}
    \caption{Ground-state energy of \nucl{O}{18}  as a function of the pairing constraint $\delta$ calculated from various many-body methods.}
    \label{fig:O19_eMax4}
\end{figure}

The singly open-shell \nucl{O}{18} constitutes the first nucleus of the present study in which static correlations are expected to be significant. In this particular case, static correlations relate to superfluidity and thus translate first at the HFB level into the spontaneous breaking of the $U(1)$ global-gauge symmetry associated with particle number conservation. Correspondingly, the pairing gap operator is used here as a constraint to vary the amount of pairing correlations in the HFB seeds~\cite{Duguet:2020hdm}. As a next step, further static correlations are captured via the restoration of neutron number and the inclusion of pairing fluctuations through the PGCM.

\begin{figure}
    \centering
    \includegraphics[width=.5\textwidth]{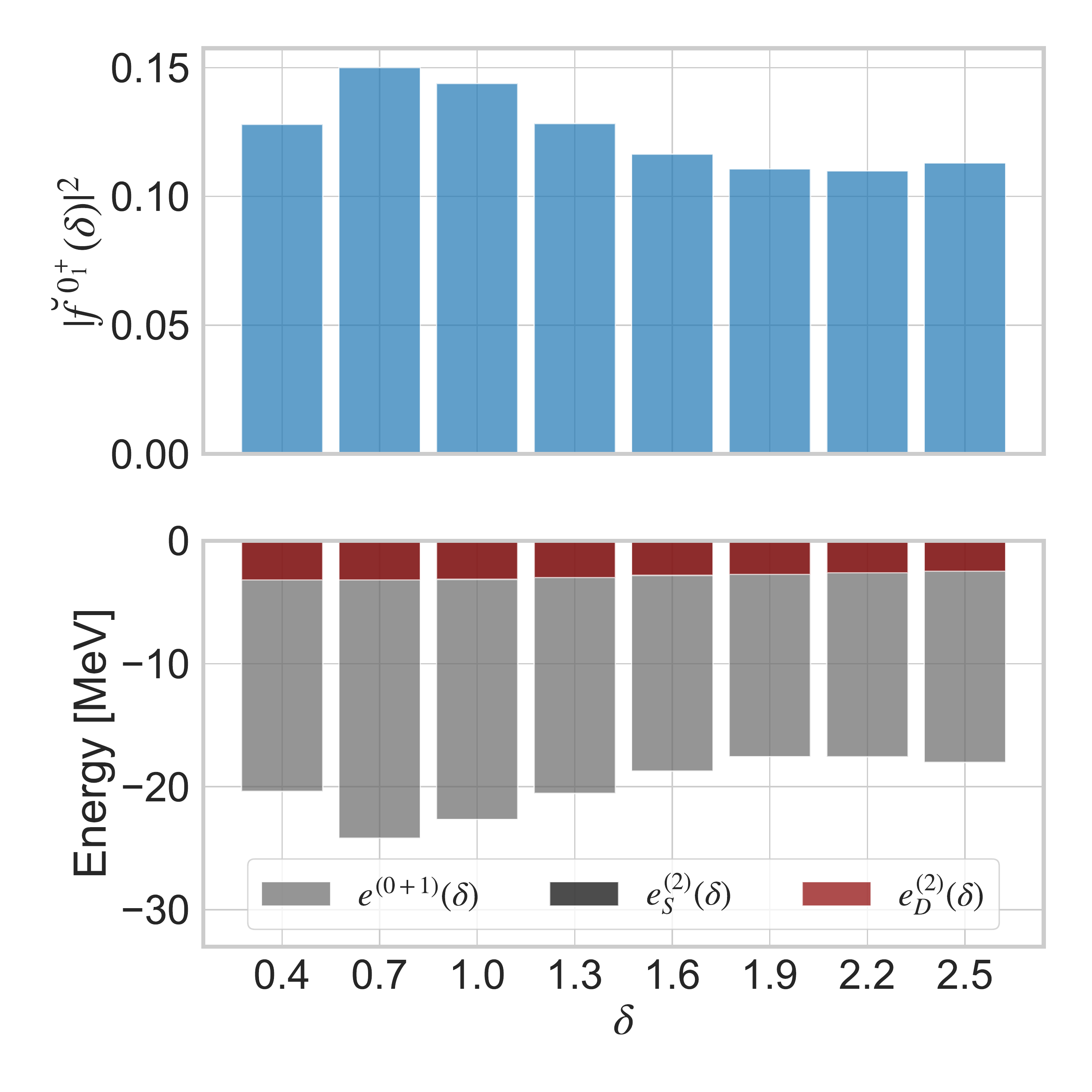}
    \caption{(Color online) Upper panel: collective ground-state PGCM wave-function  probability distribution  ($|\breve{f}^{0^+_1}(\delta)|^2$) in \nucl{O}{18} as a function of the pairing constraint $\delta$ characterizing the underlying HFB vacua. Lower panel: contributions to the PGCM ($e^{(0+1)}_{0}(\delta)$) ground-state energy and to the PGCM-PT(2) ($e^{(2)}_{S}(\delta)+e^{(2)}_{D}(\delta)$) correlation energy. The latter is split into single (two quasi-particle) and double (four quasi-particle) contributions.}
    \label{fig:O18_wave_functions}
\end{figure}

The ground-state TECs of \nucl{O}{18} are displayed in Fig.~\ref{fig:O19_eMax4} as a function of the pairing constraint $\delta$ of the (underlying) HFB vacua\footnote{See Ref.~\cite{Duguet:2020hdm} for the definition of $\delta$.}.  By definition $\delta=1$ corresponds to the canonical, i.e. unconstrained, HFB solution. While the PHFB TEC follows the HFB one, it is less bound, e.g. by $1.2$\,MeV at the canonical point. The fact that the particle-number projection after variation (PNPAV) decreases the binding reflects the fact that the distribution of particle numbers in the HFB state around the average is distorted towards heavier systems. In the next step, the GCM mixing associated with the inclusion of pairing fluctuations yields negligible correlation energy compared to the PNPAV that provides the essential IR correlations. 

Similarly to \nucl{O}{16}, PGCM underbinds the FCI result by about $25$\,MeV ($\sim13\%$), thus missing significant IR dynamical correlations. While formally not identical to canonical BMBPT(2)\footnote{See the related discussion in Paper I.}, the single-reference reduction of PGCM-PT(2), here denoted by HFB-PT(2), captures dynamical correlations on top of HFB. Results from both single-reference methods are very similar and agree with FCI within uncertainties. 

However, this close agreement is accidental and somewhat spurious. Indeed, PHFB-PT(2), which actually corrects for the $U(1)$ breaking of HFB-PT(2), pushes the energy up by about $1.5$\,MeV away from the FCI result at the canonical point. This number is close to the difference between HFB and PHFB mentioned above. While the present calculation constitutes the first example demonstrating the impact of exactly restoring symmetries within (perturbative) expansion methods, it is seen below that the $1.5$\,MeV shift shall not be fully attributed to the symmetry restoration. Adding the GCM mixing into the unperturbed state, the PGCM-PT(2) result remains consistent with PHFB-PT(2) at the canonical point within uncertainties.  

Going away from the canonical point, BMBPT(2) and HFB-PT(2) differ. This behavior reflects the different nature of the partitionings used by the two expansions, which is magnified as one departs from the canonical point. At the same time, PHFB-PT(2) becomes less (more) bound than  PGCM-PT(2) as $\delta$ becomes smaller (greater) than $1$. Once again, these behaviors do not instill trust in perturbative results away from the canonical point. Thankfully, PGCM-PT(2) is better controlled given that the configurations associated with different values of the collective coordinate $\delta$ enter the unperturbed PGCM state with weights dictated by the physical Hamiltonian. As shown in Fig.~\ref{fig:O18_wave_functions}, the collective PGCM wave-function spreads significantly on both sides of the canonical point with a maximum located to the left of it $(\delta=0.7)$. While the decomposition of the PGCM energy reflects this distribution, the second-order correction is flatter with $\delta$ but slightly favors values smaller than 1. Eventually, the PGCM-PT(2) binding energy is very close to the PHFB-PT(2) energy at the canonical point and lies $1.5$\,MeV (0.8\%) above the FCI result.

Still, PGCM-PT(2) and PHFB-PT(2) results carry error bars associated with the approximate solution of the linear system at play in the formalism. Due to to its large dimension, this linear system is solved iteratively as discussed in App.~\ref{solvingLS}, introducing an uncertainty that can be evaluated through Eq.~\eqref{estimatederror}. The solution can also be affected by linear redundancies and intruder problems that are dealt with via the simultaneous use of a norm preconditioning and a complex shift $\gamma$ as detailed in Apps.~\ref{subsec:precond} and~\ref{sec:shift}, respectively. While increasing the precision, the use of an overly large complex shift may degrade the accuracy by generating a bias in the extracted value. 

In \nucl{O}{18}, which qualifies as a difficult case, the iterative procedure can be converged in a stable fashion with a complex shift \(\gamma=10\)\,MeV, eventually leading to a $\pm 0.3$\,MeV precision on the PGCM-PT(2) energy\footnote{The precision on the PHFB-PT(2) is better ($\pm 0.1$\,MeV) thanks to the lower dimension and the near diagonal character of the linear system.} that is visualized by a band in Fig.~\ref{fig:O19_eMax4}. While the central value reported in Fig.~\ref{fig:O19_eMax4} is obtained for \(\gamma=10\)\,MeV, the bias (not reported on the figure) due to this complex shift\footnote{The bias is estimated by varying the shift over the interval $\gamma\in[5,15]$\,MeV, see App.~\ref{sec:shift_ill} for an illustration.} pushes the PGCM-PT(2) energy up by about $1$\,MeV. Eventually, the bias accounts for two thirds of the $1.5$\,MeV (0.8\%) disagreement with the FCI result and for two thirds of the shift upward compared to HFB-PT(2) that was  fully attributed to the symmetry restoration at first.

\begin{figure}
    \centering
    \includegraphics[width=.5\textwidth]{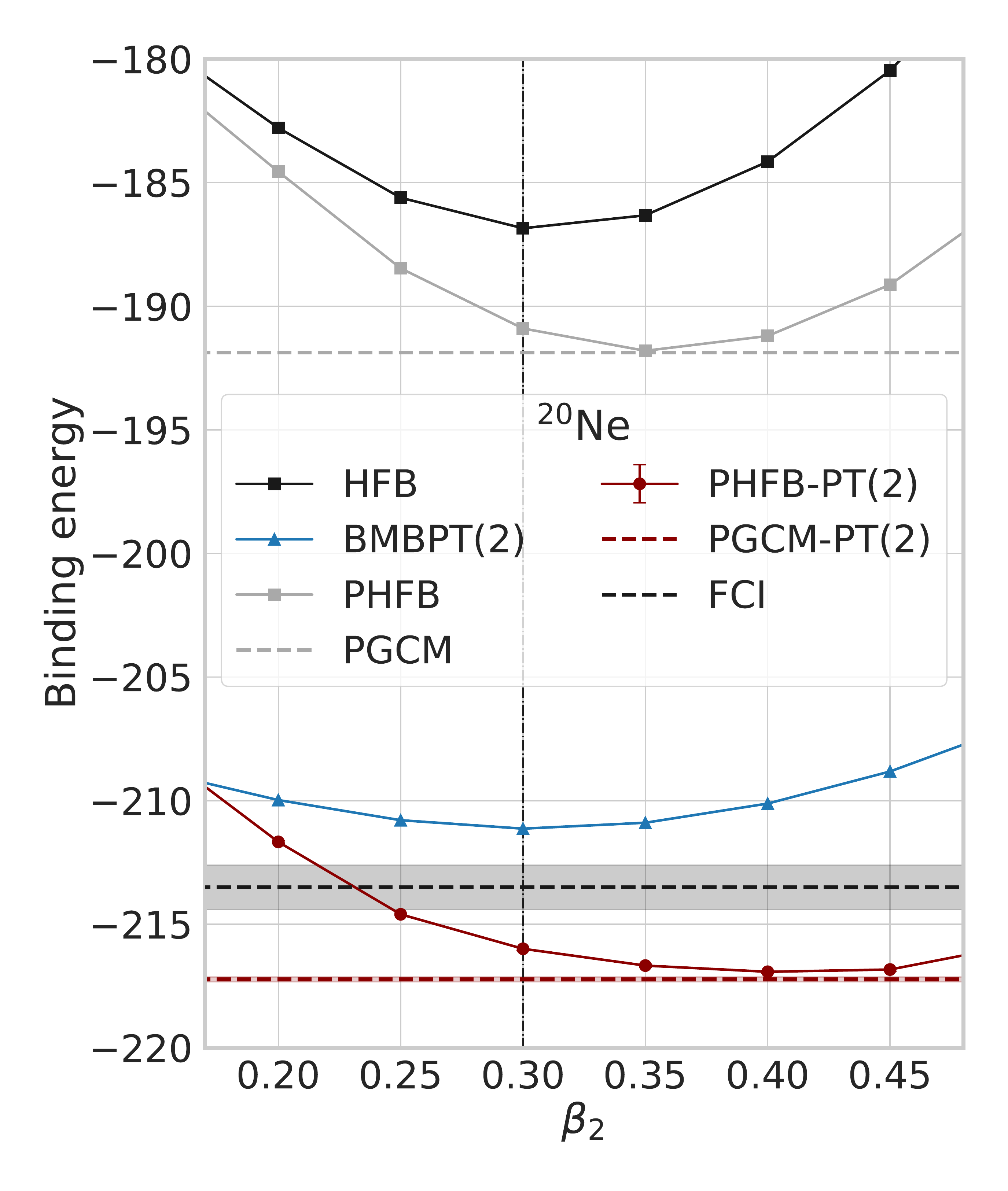}
    \caption{Ground-state energy of \nucl{Ne}{20} as a function of the axial quadrupole deformation $\beta_2$ of the (underlying) HFB states.}
    \label{fig:Ne20_eMax4}
\end{figure}

\subsubsection{\nucl{Ne}{20}}
\label{SecNe20emax4}

The doubly open-shell \nucl{Ne}{20} displays strong static correlations that manifest through the breaking of $SU(2)$ rotational symmetry associated with angular momentum conservation at the HFB level. Accordingly, the axial quadrupole moment operator is used as a constraint to vary the deformation of the HFB seeds. As a next step, further static correlations are captured via the restoration of angular momentum and the inclusion of shape fluctuations through the PGCM. As demonstrated in Paper II, the description of \nucl{Ne}{20} strongly benefits from breaking and restoring parity as well as the inclusion of octupole shape fluctuations. Our present calculations are however restricted to axial quadrupole deformation, leaving some room for further improvement in the future. While $U(1)$ global gauge symmetry is also allowed to break spontaneously, it does not do so with the presently employed Hamiltonian, hence all HFB states actually reduce to (deformed) HF Slater determinants.

\begin{figure}
    \centering
    \includegraphics[width=.5\textwidth]{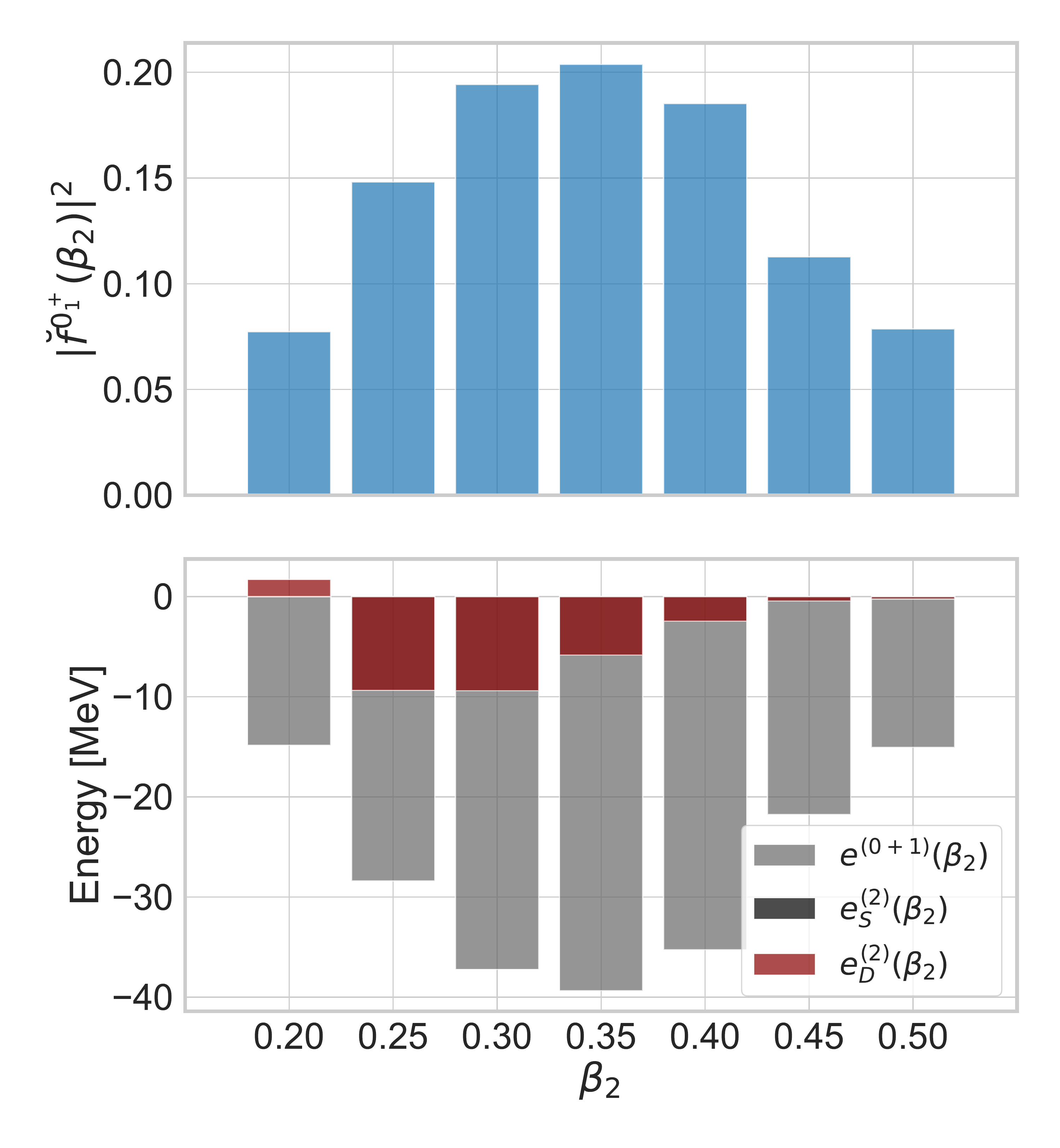}
    \caption{(Color online) Upper panel: collective ground-state PGCM wave-function probability distribution  ($|\breve{f}^{0^+_1}(\beta_2)|^2$) in \nucl{Ne}{20} as a function of the axial quadrupole deformation $(\beta_2)$ of the underlying HFB vacua. Lower panel: contributions to PGCM ($e^{(0+1)}_{0}(\beta_2)$) and PGCM-PT(2) ($e^{(2)}_{S}(\beta_2)+e^{(2)}_{D}(\beta_2)$) ground-state energies as a function of the axial quadrupole deformation $(\beta_2)$ of the underlying HFB vacua. The PGCM-PT(2) constributions are split into singles (two quasi-particle) and doubles (four quasi-particle) contributions.}
    \label{fig:Ne20_wave_functions}
\end{figure}

\begin{figure}
    \centering
    \includegraphics[width=.5\textwidth]{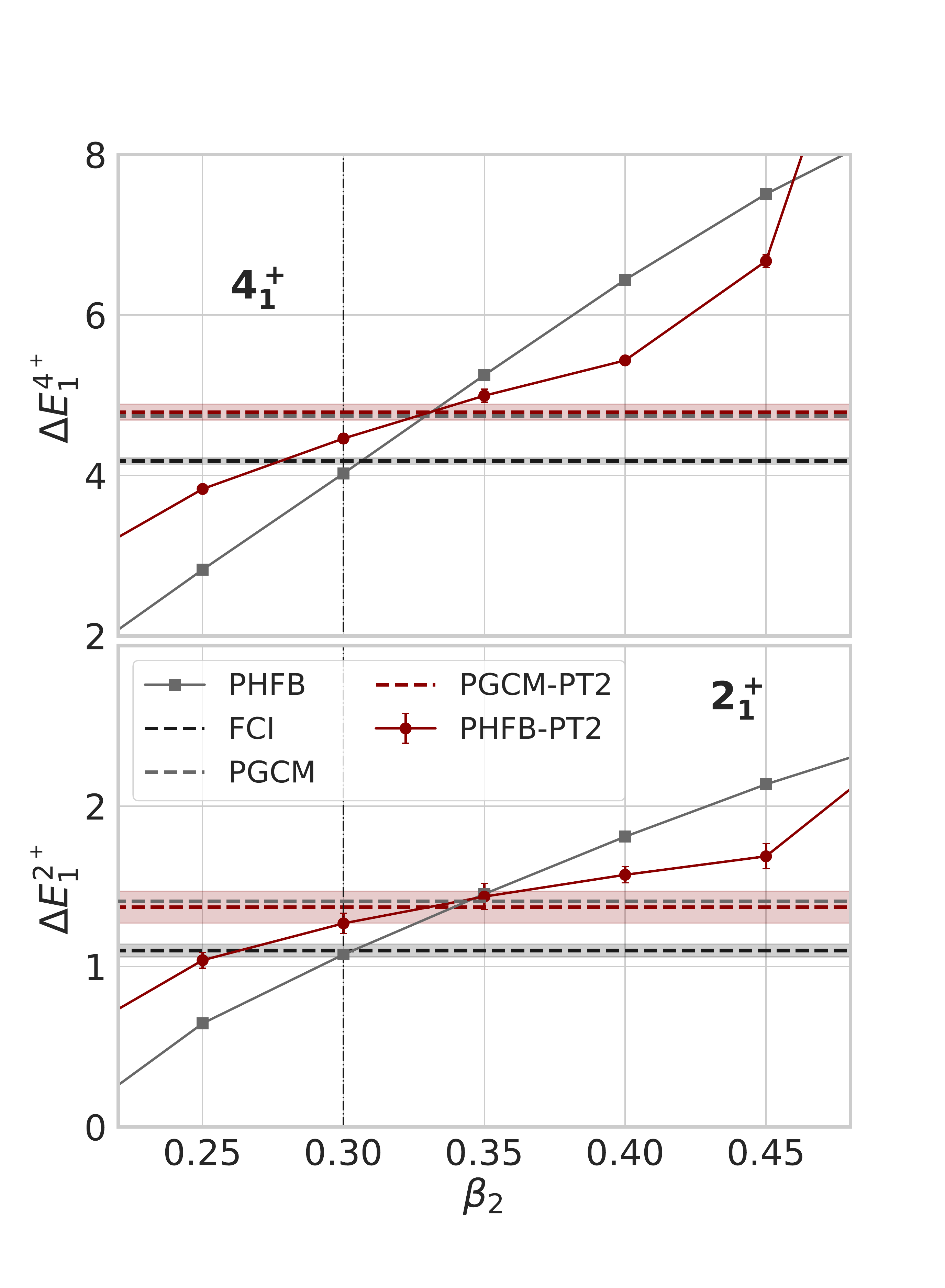}
    \caption{(Color online) Excitation energy in \nucl{Ne}{20} as a function of the axial quadrupole deformation $(\beta_2)$ of the underlying HFB vacua. Top panel: first \(4^+\) state. Bottom panel: first \(2^+\) state. Calculations are performed with $\hbar \omega=20\text{ MeV}$, $e_{\text{max}}=4$ and employing the two-body part of the N$^3$LO $\chi$EFT Hamiltonian evolved to $\lambda_{\text{vsrg}}=1.88 $\,fm$^{-1}$.}
    \label{fig:Ne20_eMax4_exc}
\end{figure}

The ground-state TECs of \nucl{Ne}{20} are displayed in Fig.~\ref{fig:Ne20_eMax4} as a function of the axial quadrupole deformation\footnote{See Paper II for a precise definition.} $\beta_2$ of the (underlying) HFB vacua. One first observes that the projection on $J$ provides a significant energy gain of $5.5$\,MeV and moves the minimum of the PHFB TEC to larger deformation ($\beta_2=0.35$) than the canonical HFB minimum ($\beta_2=0.3$). The GCM mixing only adds $80$\,keV correlation energy given that the TEC is rather stiff along the axial quadrupole direction\footnote{As shown in Paper II, the energy is softer against axial octupole deformations.}. Once again, static correlations are dominated by the symmetry restoration. Having included essential static correlations, the PGCM energy is still $21.7$\,MeV ($10\%$) away from the FCI result, and misses significant dynamical correlations. 

Stepping back to canonical HFB and adding dynamical correlations via  BMBPT(2) lowers the energy by $24.6$\,MeV, yielding a result that is $2.6$\,MeV ($1.2\%$) underbound compared to FCI\footnote{Canonical BMBPT(2) is the closest point to FCI along the TEC in the present example. Note that canonical BMBPT(3) only provides an extra $0.3$\,MeV correlation energy compared to canonical BMBPT(2).}. 

On the other hand, starting from the PHFB TEC and adding dynamical correlations via PHFB-PT(2) lowers the energy by $25.1, 24.9$ and $25.7$\,MeV at the HFB, PHFB, and PHFB-PT(2) minima, respectively. These energies overshoot the FCI result by about $2.5/3.2/3.4$\,MeV ($1.2/1.5/1.6\%$). While the difference between BMBPT(2) and PHFB-PT(2) TECs is similar to the difference between HFB and PHFB TECs, one observes that a consistent angular-momentum restoration favors larger deformations when adding dynamical correlations. 

The mixing of quadrupole shapes in PGCM-PT(2) only adds $310$\,keV to the PHFB-PT(2) minimum. The PGCM-PT(2) result keeps a close memory of the PHFB-PT(2) minimum ($\beta_2=0.4$) rather than the PHFB-PT(2) value at the canonical HFB minimum ($\beta_2=0.3$). All in all, the PGCM-PT(2) energy\footnote{Present PGCM-PT(2) and PHFB-PT(2) results were obtained with a complex shift $\gamma=15$\,MeV. The precision error associated with solving the linear system is shown through an error band in Fig.~\ref{fig:Ne20_shift_4}.} overshoots the FCI result by \(1.7\%\). This discrepancy is expected to decrease after the inclusion of the octupole degree of freedom into the PGCM. 

\begin{figure}
    \centering
    \includegraphics[width=.5\textwidth]{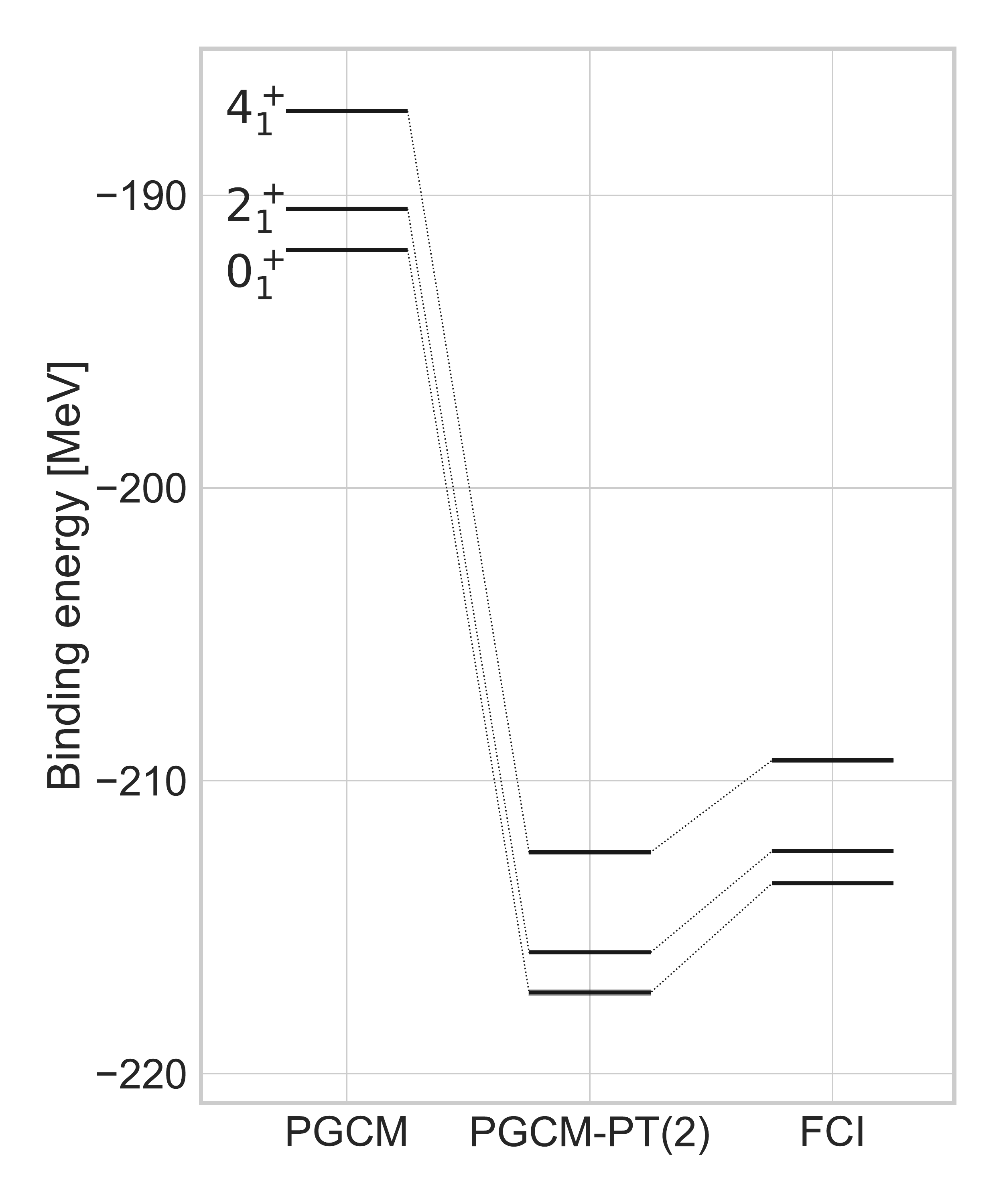}
    \caption{Absolute energies of the first $0^+$, $2^+$ and $4^+$ states in \nucl{Ne}{20} computed via PGCM, PGCM-PT(2) and FCI.}
    \label{fig:Ne20_eMax4_spectrum}
\end{figure}

In order to further analyse the theoretical content of the above results,  Fig.~\ref{fig:Ne20_wave_functions} shows that the collective PGCM ground-state wave-function and the associated energy contributions are distributed rather symmetrically around the $J^\pi = 0^+$ PHFB minimum ($\beta_2=0.35$) of the TEC visible in Fig.~\ref{fig:Ne20_eMax4} and spread over a large interval of $\beta_2$ values. Interestingly, dynamical correlations captured via PGCM-PT(2) favor configurations\footnote{Once again, single excitations bring negligible contributions to the correlation energy.} to the left of the HFB minimum ($\beta_2=[0.25,0.30]$). As a result, dynamical correlations could counterbalance the overestimated radii obtained at the PGCM level (see Paper II) due to the opposite predilection of the latter for deformations larger than the HFB minimum. This interesting and non-trivial finding will have to be confirmed by an explicit calculation of rms radii at the PGCM-PT(2) level in the future.

In addition to providing accurate absolute energies in complex systems, e.g. in doubly open-shell nuclei displaying strong collective static correlations, a key advantage of the multi-reference PGCM-PT formalism over BMBPT is that it provides natural access to the low-lying spectroscopy within a symmetry-conserving scheme by correcting each PGCM eigenstate for dynamical correlations.

The first \(2^+\) and \(4^+\) excitation energies in \nucl{Ne}{20}  are shown in Fig.~\ref{fig:Ne20_eMax4_exc} as a function of the axial quadrupole deformation. First, one observes that the PGCM \(2^+_1\) and \(4^+_1\)  excitation energies differ from the FCI results by $300$\,keV ($27\%$) and $560$\,keV ($13\%$), respectively. This is consistent with the results displayed in Paper II. One also sees that PHFB results at the canonical deformation  ($\beta_2=0.3$) are very close to PGCM ones, but the differences grow for smaller or large deformations. Adding dynamical correlations, PHFB-PT(2) flattens the excitation energies as a function of $\beta_2$ compared to PHFB, systematically going into the direction of PGCM-PT(2) for each deformation. Given that exact results would be independent of the deformation of the underlying vacuum, this feature is an empirical sign that PHFB-PT(2) results are better converged than PHFB ones. It also implies that the PGCM-PT(2) spectrum converges with fewer states than the PGCM one.  Still, at the canonical deformation  ($\beta_2=0.3$) dynamical correlations are small, which remains true even when shape mixing is added, given that PGCM-PT(2) excitation energies are essentially identical to PGCM ones.

Overall, the PGCM-PT(2) \(2^+_1\) and \(4^+_1\) excitation energies differ by $24\%$ and $15\%$ from FCI results respectively, which seems to indicate that missing correlations are beyond two-particle/two-hole excitations of axially deformed HF states. While going to PGCM-PT(3) will help reduce this difference, it might be numerically less costly and more relevant in this case to enrich the PGCM unperturbed state via, e.g., the inclusion of octupole, triaxial and/or pairing degrees of freedom, or to start from HFB states obtained via a variation after particle-number-projection (VAPNP) calculation, in order to compress the spectrum. In the future, another possibility would be to design a non-perturbative extension of the multi-reference PGCM-PT formalism to more efficiently capture higher-rank particle-hole excitations.

Our \nucl{Ne}{20} results are summarized in Fig.~\ref{fig:Ne20_eMax4_spectrum} where the combined benefits of PGCM-PT are clearly apparent. Although a slight overbinding of about $3$\,MeV ($\sim1.5\%$) is observed, PGCM-PT(2) brings down absolute energies to the right range of values without degrading their relative position. 
This latter feature is far from trivial given that the PGCM-PT formalism is {\it state specific}, i.e. calculations are performed separately on top of each PGCM eigenstate, and considering that each PGCM energy is corrected by about $25$\,MeV while their relative distance is on the MeV scale. 
In particular, the (non-trivial) numerical techniques used to solve the PGCM-PT(2) equations must be well controlled to maintain the consistency of the spectra. For example, it is essential to use the same complex shift $\gamma$ for all states belonging to a given nucleus in order for the bias on absolute energies to be consistent and to largely cancel out in the excitation spectrum.

\section{Adding the MR-IMSRG  pre-processing}
\label{IMSRG_H}

In the present part, PGCM-PT(2) calculations are performed in a larger model space with \(e_{\text{max}}=6\) (and $\hbar\omega=16$\,MeV)\footnote{The limitation to \(e_{\text{max}}=4\) was due to the wish to benchmark PGCM-PT(2) calculations against FCI results.}. We use a Hamiltonian consisting of an SRG-evolved chiral N${}^{3}$LO nucleon-nucleon interaction with \(\lambda_{\text{vsrg}}=1.8\text{ fm}^{-1}\), supplemented with an N${}^{2}$LO three-nucleon interaction with cutoff $\Lambda=2.0\,\mathrm{fm}^{-1}$ whose low-energy constants are adjusted to $A=3,4$ observables, as described in Refs.~\cite{PhysRevC.83.031301,Nogga:2004il}. The Hamiltonian is further pre-processed via the MR-IMSRG unitary transformation based on the $J^\pi=0^+$ canonical PHFB state. The evolutions are based on the MR-IMSRG(2) truncation scheme, employing the so-called Brillouin generator --- see Refs.~\cite{Hergert:2015awm,Hergert:2016etg} for details. The MR-IMSRG transformation is parametrized by the flow parameter \(s\in[0,20]\, \mathrm{MeV}^{-1}\), where $s=0$ means that no transformation is applied and the upper limit is chosen such that the transformed Hamiltonian no longer exhibits significant evolution. In the MR-IMSRG(2) truncation scheme, the three-nucleon interaction included at the beginning of the flow (along with higher-body operators for $s>0$) is approximated via the normal ordering with respect to the J=0 PHFB state. 

In closed-shell nuclei (not shown here), the PHFB reference state reduces to a spherically invariant Slater determinant such that MR-IMSRG is nothing but the simpler SR-IMSRG method. In this case, pushing the transformation to \(s=\infty\) ($s$ sufficiently large in practice) leads to a complete resummation of dynamical correlations into the pre-processed Hamiltonian such that the unperturbed HF Slater determinant becomes its exact ground state, i.e. no further correlations need to be added. While dynamical correlations are largely resummed in open-shell nuclei via the MR-IMSRG pre-processing, the decoupling of the reference state cannot be complete~\cite{Hergert:2016etg} such that an additional step is always needed to grasp the remaining correlations as illustrated below\footnote{While the exact decoupling is formally realized in the simpler SR-IMSRG method, it can usually be achieved to good accuracy in actual MR-IMSRG calculations when pushing the transformation far enough. Strictly speaking, however, the decoupling cannot be complete in MR-IMSRG, at least for generators that have an explicit and manageable second-quantized representation. One may be able to define a generator that formally achieves the decoupling but such a generator would have no practical, i.e. low-rank, representation.}.

\subsection{\nucl{O}{18}}

\begin{figure}
    \centering
    \includegraphics[width=.5\textwidth]{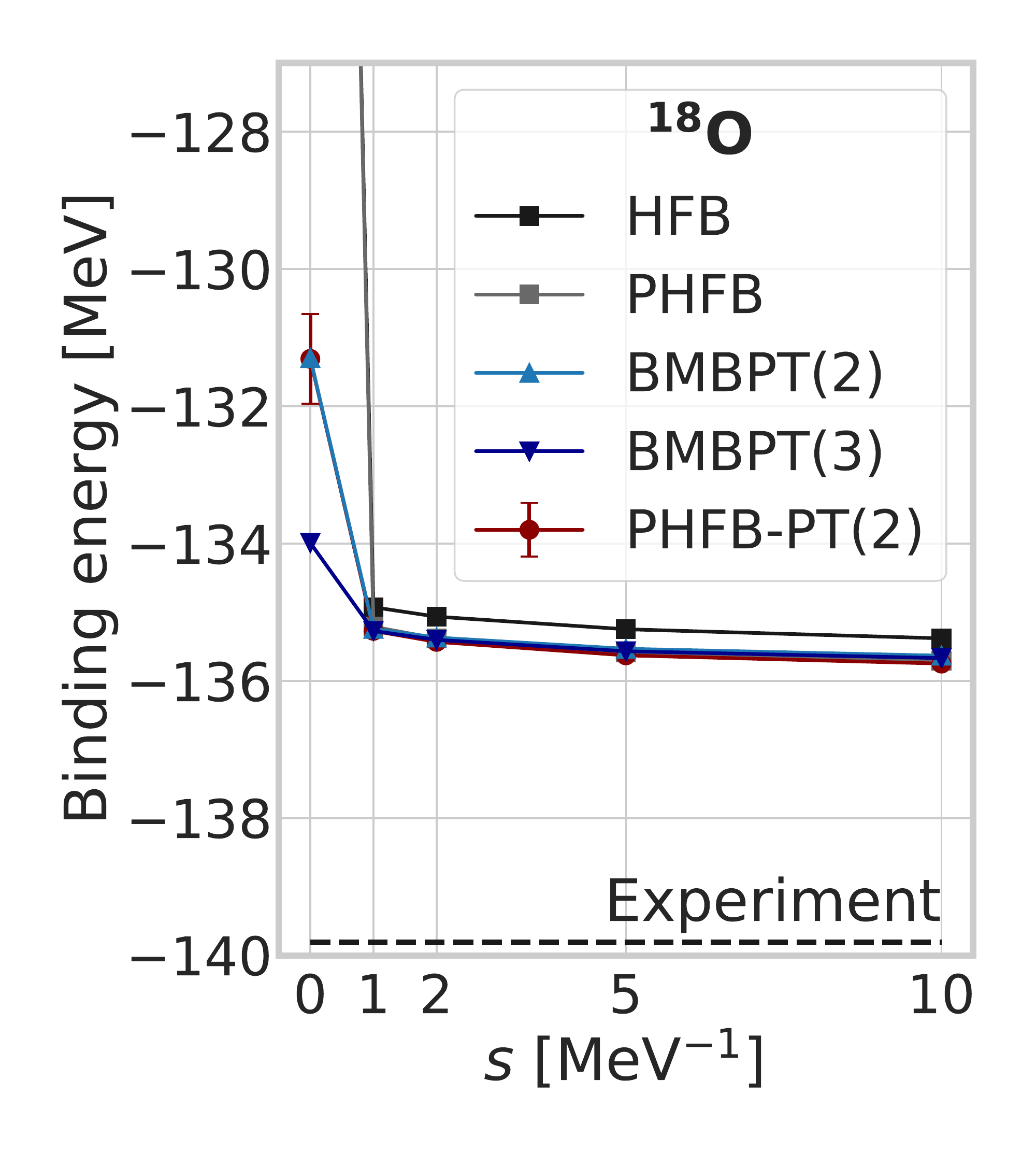}
    \caption{(Color online) Absolute binding energy of \nucl{O}{18} as a function of the flow parameter \(s\) associated with the MR-IMSRG pre-processing of the Hamiltonian.}
    \label{fig:O18_MSU_gs}
\end{figure}

The absolute binding energy of \nucl{O}{18} is displayed in Fig.~\ref{fig:O18_MSU_gs} as a function of $s$. Due to the PNPAV in \nucl{O}{18}, the number of single and double excitations of the HFB vacuum required to perform a PHFB-PT(2) calculation is already very large ($10^6$ states) for \(e_{\text{max}}=6\). The numerical implementation will be optimized in the future, but in the mean-time, the calculation is made faster by discarding configurations based on their norm as specified in App.~\ref{App_numerics}. For the same reason, only PHFB-PT(2) calculations on top of the spherical \nucl{O}{18} canonical HFB vacuum have been performed, leaving a PGCM-PT(2) calculation for the future.

\begin{figure}
    \centering
    \includegraphics[width=.5\textwidth]{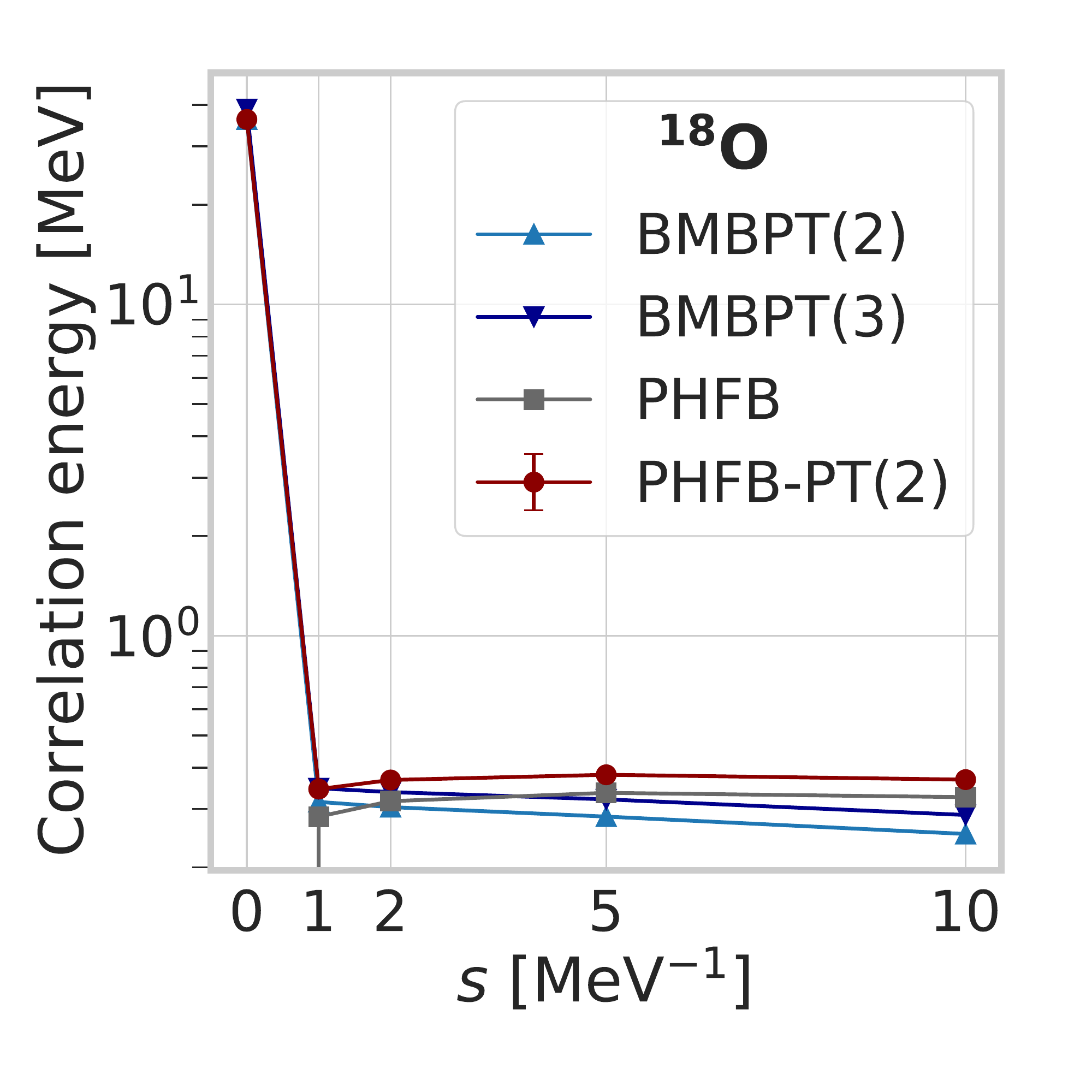}
    \caption{(Color online) Correlation energy, i.e. difference to the canonical HFB result, in \nucl{O}{18} as a function of the flow parameter \(s\) associated with the MR-IMSRG pre-processing of the Hamiltonian.}
    \label{fig:O18_MSU_corr}
\end{figure}

In spite of the change of model space and Hamiltonian, the situation encountered at $s=0$  is qualitatively similar to the one discussed in Sec.~\ref{VSRG_H}. Indeed, while HFB and PHFB are largely underbound, BMBPT(2) and PHFB-PT(2) bring in the dominant fraction of dynamical correlations\footnote{Contrarily to the results obtained in Sec.~\ref{O182body} with a two-body interaction only and \(e_{\text{max}}=4\), PHFB-PT(2) is very close to BMBPT(2) at $s=0$. At the same time, the contribution of BMBPT(3) is enlarged.}, with BMBPT(3) adding an extra $2$\,MeV. Switching on the MR-IMSRG pre-processing, HFB and PHFB energies drop dramatically for small values of $s$ and flatten out very quickly beyond $s=1\,\mathrm{MeV}^{-1}$. At the same time, BMBPT(2), PHFB-PT(2) and BMBPT(3) drop towards a similar value, about $1.5$\,MeV below the original BMBPT(3) result, which happens to be also similar to the PHFB value. Eventually, PHFB-PT(2) is about $4$\,MeV ($2.9\%$) away from experiment. No convergence analysis as a function of the model space has been performed and reaching a converged  absolute binding energy clearly requires (an extrapolation to) a larger model space.

To better appreciate the impact of the MR-IMSRG evolution, the correlation energy, i.e. the difference to the HFB result, is shown in Fig.~\ref{fig:O18_MSU_corr}. Having already absorbed the bulk of dynamical correlations, pre-processed Hamiltonians become more and more perturbative with increasing $s$ such that BMBPT(2,3) and PHFB-PT(2) corrections become less important with the flow, i.e. one goes from $38.8$\,MeV and $36.2$\,MeV for BMBPT(3) and PHFB-PT(2) at $s=0$ to $288$\,keV and $369$\,keV at $s=10\,\mathrm{MeV}^{-1}$, respectively, with an inversion of the two results. At the same time, the particle number projection that is repulsive at $s=0$ ($-394$\,keV) brings in additional binding for $s\geq 1\,\mathrm{MeV}^{-1}$ ($+327$\,keV at $s=10\,\mathrm{MeV}^{-1}$). These results demonstrate that correlations are reshuffled through the MR-IMSRG flow, such that the importance of dynamical correlations is strongly reduced whereas static correlations are somewhat enhanced. 

Dynamical correlations added on top of PHFB via PHFB-PT(2)\footnote{The numerical solution of the PHFB-PT(2) linear system is very stable in the present example such that a small complex shift ($\gamma=1$\,MeV) can be used safely. The precision error on PHFB-PT(2) energies  is essentially invisible in Fig.~\ref{fig:O18_MSU_corr} whereas the bias generated for $\gamma=1$\,MeV is negligible compared to the $42$\,keV difference between PHFB and PHFB-PT(2) energies at $s=10\,\mathrm{MeV}^{-1}$.} become as small as $42$\,keV at $s=10\,\mathrm{MeV}^{-1}$. Thus, the PHFB state used as a reference for the MR-IMSRG pre-processing is, for all practical purposes, decoupled from the ${\cal Q}$ space at the end of the transformation in the present calculation. Although the decoupling cannot be exact in principle, \nucl{O}{18} behaves similarly to a closed-shell nucleus such that the dynamical correlations left to be captured after PNPAV are very small.

\begin{figure}
    \centering
    \includegraphics[width=.5\textwidth]{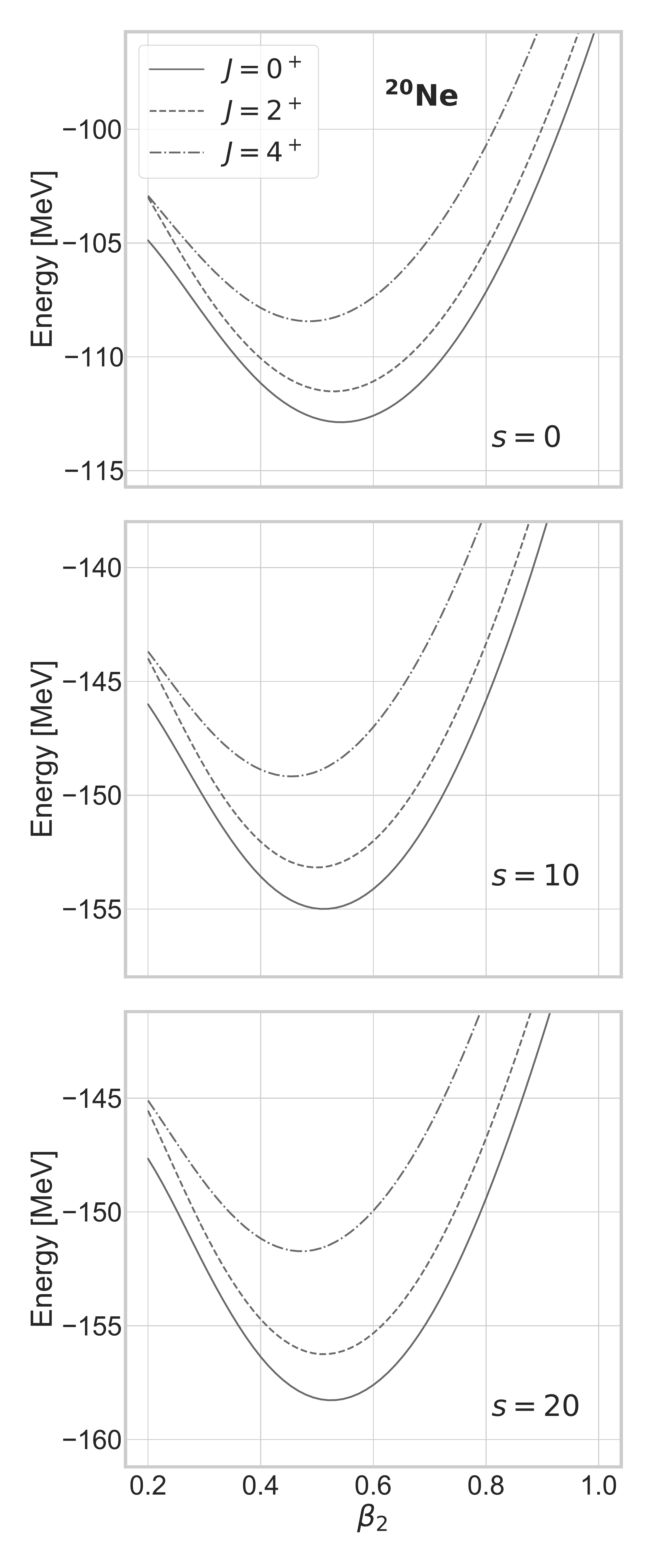}
    \caption{(Color online) $J^\pi = 0^+,2^+,4^+$ PHFB TECs in \nucl{Ne}{20}  as a function of the axial quadrupole deformation $\beta_2$ for $s=0\,\mathrm{MeV}^{-1}$ (upper panel), $s=10\,\mathrm{MeV}^{-1}$ (middle panel) and $s=20\,\mathrm{MeV}^{-1}$ (lower panel).}
    \label{fig:Ne20_TECs}
\end{figure}

\begin{figure}
    \centering
    \includegraphics[width=.5\textwidth]{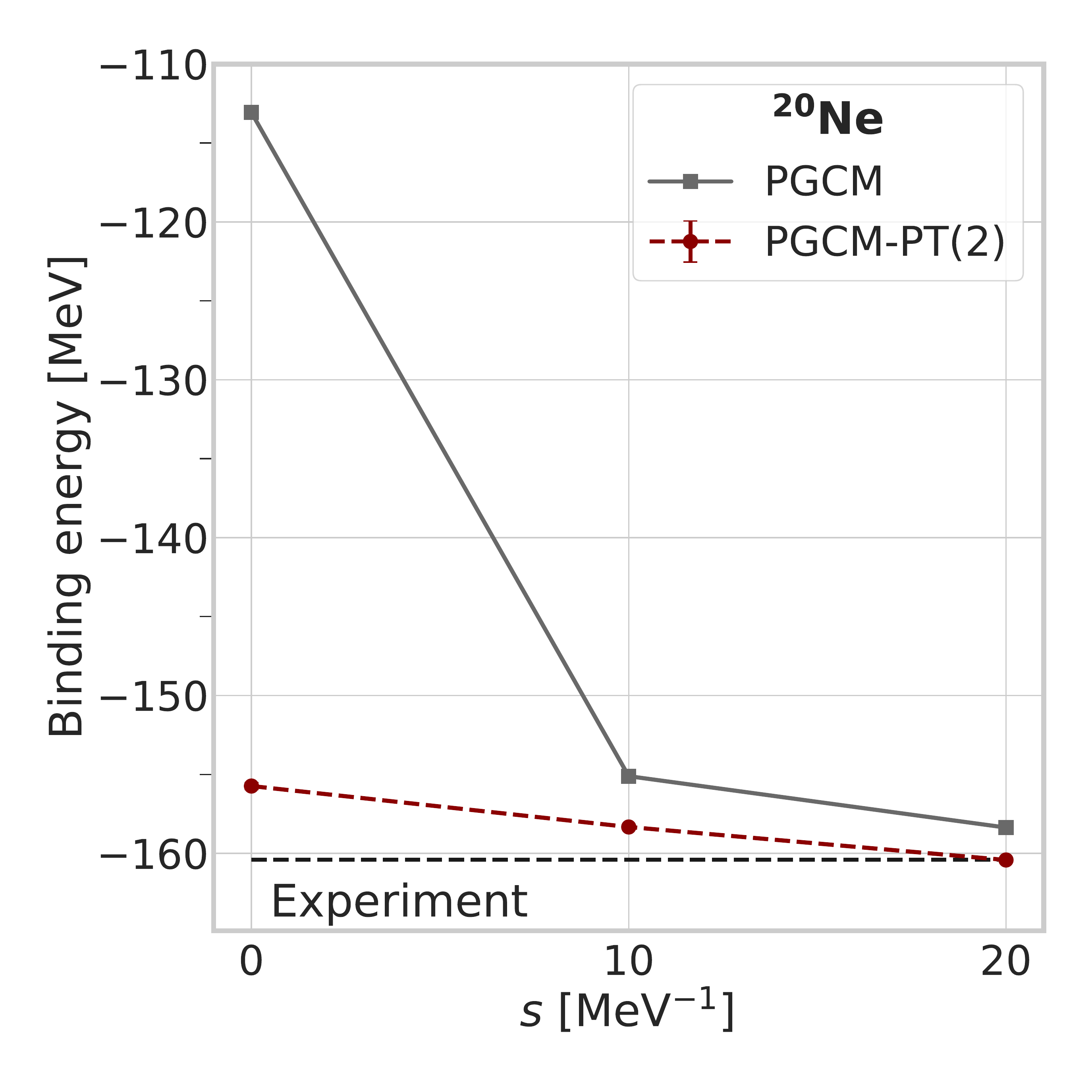}
    \caption{(Color online) Absolute PGCM and PGCM-PT(2) binding energies of \nucl{Ne}{20} as a function of the MR-IMSRG flow parameter \(s\).}
    \label{fig:Ne20_gs}
\end{figure}

\subsection{\nucl{Ne}{20}}

The doubly open-shell \nucl{Ne}{20} constitutes a richer and more instructive example.  Figure~\ref{fig:Ne20_TECs} shows the $J^\pi = 0^+,2^+$ and $4^+$ PHFB TECs as a function of the axial quadrupole deformation $\beta_2$ for three values $(s=0,10,20)\,\mathrm{MeV}^{-1}$ of the MR-IMSRG flow parameter \cite{Hergert:2016etg}. The TECs are strongly lowered with $s$, e.g. the PHFB minimum gains $45.4$\,MeV going from $s=0$ to $s=20\,\mathrm{MeV}^{-1}$, with most of the effect occuring for $0\leq s \leq 10\,\mathrm{MeV}^{-1}$. At the same time, the deformation of the PHFB minimum is lowered from $\beta_2=0.55$ to $\beta_2=0.52$ while the TECs become stiffer. 

In Fig.~\ref{fig:Ne20_gs}, PGCM and PGCM-PT(2) binding energies are displayed as a function of the flow parameter. Starting from $J^\pi = 0^+$ PHFB TECs, PGCM and PGCM-PT(2) calculations mix five HFB configurations with axial quadrupole deformations \(\beta_2=(0.3,0.4,0.5,0.6,0.7)\). Unlike in \nucl{O}{18}, the convergence of PGCM energies is not fully reached yet for $s=20\,\mathrm{MeV}^{-1}$. Still, the bulk of dynamical correlations has already been resummed into the pre-processed Hamiltonian at \(s=10\), which suggests a convergent behavior. Eventually, the PGCM energy is lowered by $45.2$\,MeV between $s=0$ and $s=20\,\mathrm{MeV}^{-1}$. At the same time, PGCM-PT(2)  systematically lowers the PGCM value, the added dynamical correlations reducing from $42.5$\,MeV at $s=0$ to only $2.0$\,MeV at $s=20\,\mathrm{MeV}^{-1}$. Similarly, the difference between PHFB and PGCM-PT(2) is drastically reduced as $s$ grows but does not vanish, i.e. it still amounts to $2.03$\,MeV with the most pre-processed Hamiltonian\footnote{PHFB and PGCM energies differ by less than $200$\,keV all throughout the interval $s\in[0,20]\,\mathrm{MeV}^{-1}$.}. This indicates that, while very effective, the decoupling of the PHFB state from the ${\cal Q}$ space is not complete and thus less effective than in the singly open-shell \nucl{O}{18}. This feature points to the stronger multi-reference character of \nucl{Ne}{20} associated with the breaking and restoration of $SU(2)$ symmetry.

\begin{figure*}
    \centering
    \includegraphics[width=\textwidth]{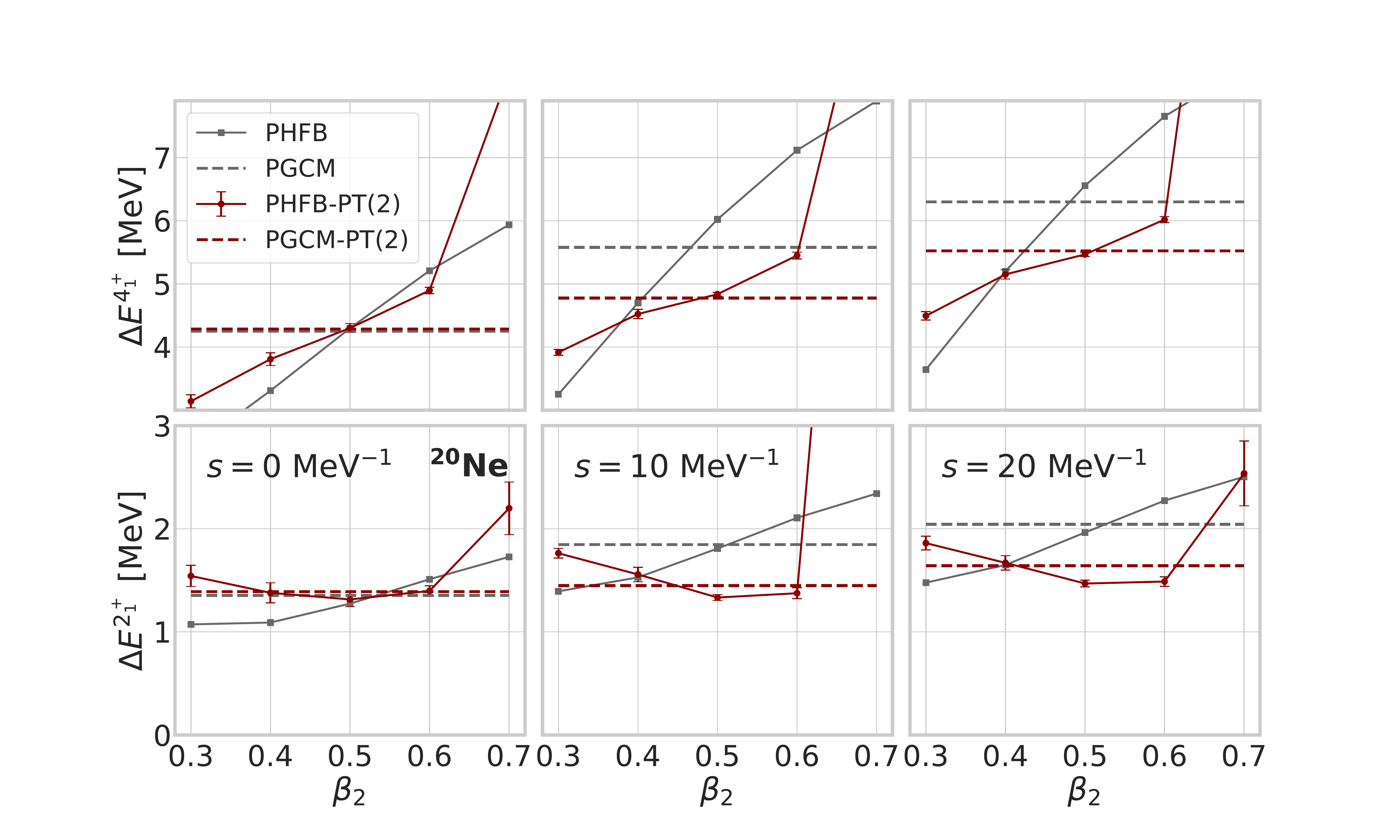}
    \caption{(Color online) \(2_1^+\) (bottom row) and \(4_1^+\) (top row) excitation energies as a function of $\beta_2$ for $s=0$ (left column), $s=10\,\mathrm{MeV}^{-1}$ (middle column) and $s=20\,\mathrm{MeV}^{-1}$ (right column).}
    \label{fig:Ne20_MSU_2+4+}
\end{figure*}

The  PGCM-PT(2) energy changes by less than $5$\,MeV over the interval $s\in[0,20]\,\mathrm{MeV}^{-1}$, thus strongly reducing the flow parameter dependence compared to PGCM results. The residual dependence of the ground-state energy on the flow parameter results both from the breaking of unitarity associated with the truncation of the flow equations at the MR-IMSRG(2) level and from the approximations to the solution of the $A$-body Schr{\"o}dinger's equation at the PGCM-PT(2) level. Under the hypothesis that the PGCM-PT is convergent and given that the second-order correction reduces to $2$\,MeV at $s=20\,\mathrm{MeV}^{-1}$, one can speculate that the PGCM-PT(2) energy is eventually better converged than the $5$\,MeV spread over the interval $s\in[0,20]\,\mathrm{MeV}^{-1}$, i.e. by better than $3\%$. The fact that the experimental value is consistent with the  PGCM-PT(2) prediction within estimated uncertainty must not be overinterpreted given that improving over the presently used $e_{\text{max}}$ truncation is expected to lower the ground-state energy by several MeVs.

\begin{figure}
    \centering
    \includegraphics[width=.5\textwidth]{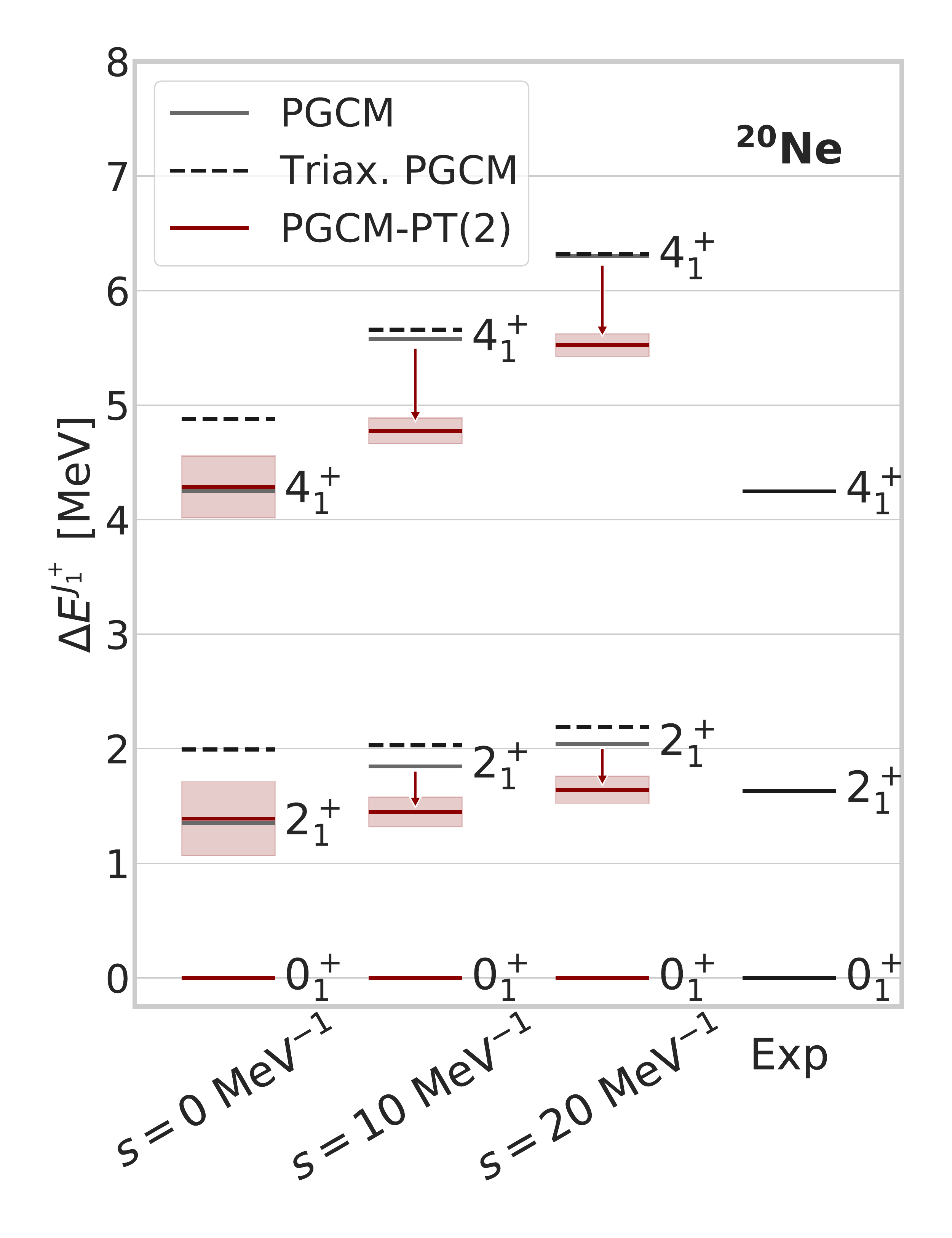}
    \caption{(Color online) Low lying spectrum of \nucl{Ne}{20} as a function of the MR-IMSRG flow parameter.}
    \label{fig:Ne20_MSU_spectrum}
\end{figure}

Turning to the low-lying spectroscopy, Fig.~\ref{fig:Ne20_MSU_2+4+} displays the first $2^+$ and $4^+$ excitation energies as a function of $\beta_2$ for the three values of the flow parameter. Focusing first on \(s=0\), the conclusions drawn in Sec.~\ref{SecNe20emax4} remain valid, i.e. PHFB-PT(2) flattens the excitation energies as a function of $\beta_2$ compared to PHFB whereas dynamical correlations brought in through PGCM-PT(2) do not modify the low-lying part of the PGCM ground-state rotational band. However, the picture changes drastically when pre-processing the Hamiltonian via MR-IMSRG. Indeed, the PGCM spectrum becomes more dilated with increasing $s$. This feature is already visible from the $J^{\pi}=0^+, 2^+, 4^+$ PHFB TECs displayed in Fig.~\ref{fig:Ne20_TECs} that become more distant with increasing $s$. Based on this trend, one observes that PHFB-PT(2), while always flattening the dependence on $\beta_2$, systematically corrects for this dilatation of the rotational spectrum. This non-trivial feature is confirmed at the PGCM-PT(2) level.

\begin{figure}
    \centering
    \includegraphics[width=.5\textwidth]{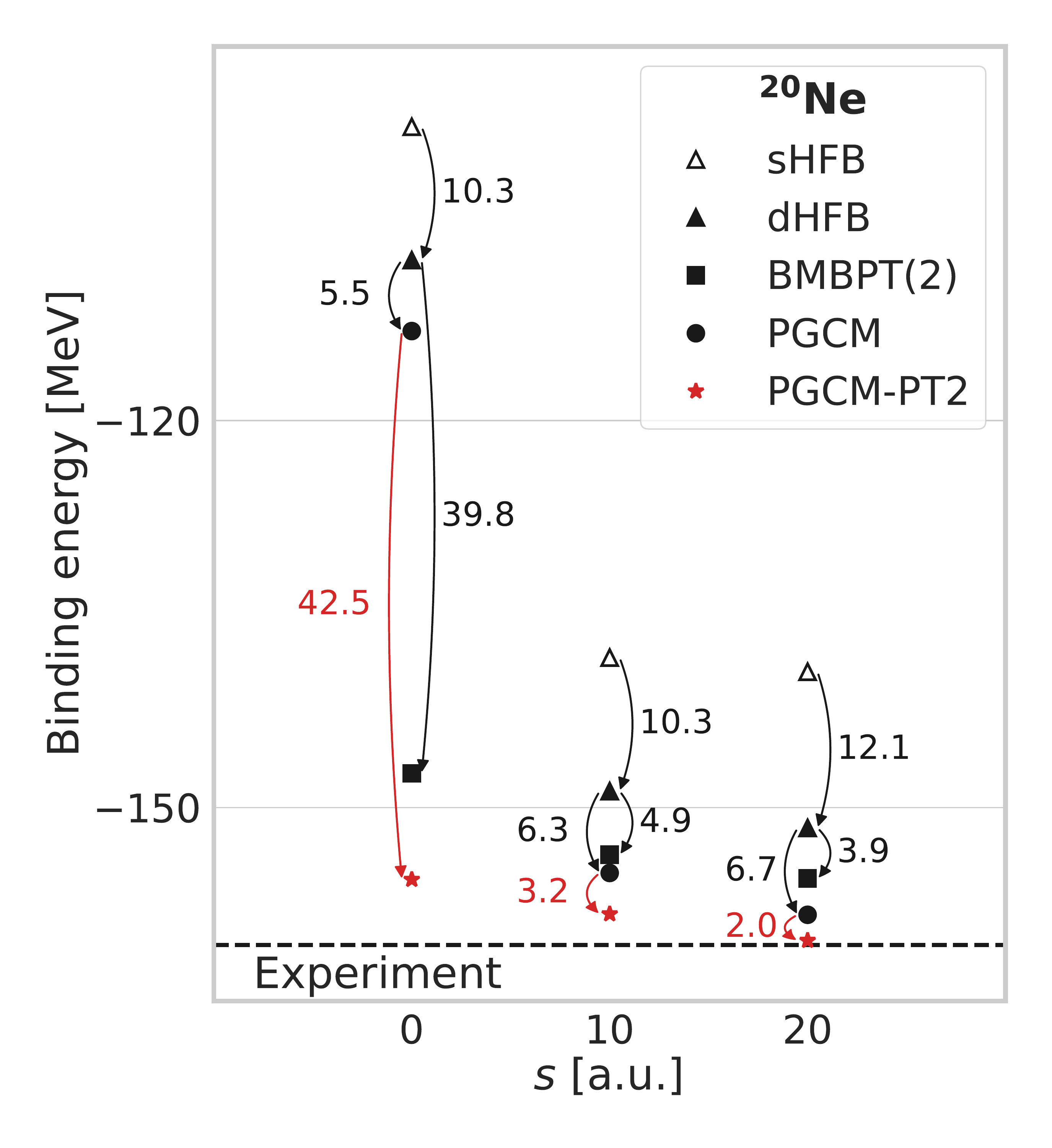}
    \caption{(Color online) Total ground-state energy of $^{20}$Ne computed within various many-body methods for three different values of the MR-IMSRG flow parameter. Numbers next to downward arrows denote the corresponding gain in correlation energy (in MeV).}
    \label{fig:Ne20_ebranplot}
\end{figure}

This key result can be better appreciated in Fig.~\ref{fig:Ne20_MSU_spectrum} where PGCM and PGCM-PT(2) spectra are compared to experiment and to the spectrum obtained from a richer PGCM calculation including additional axial states along with triaxially deformed ones. Although the PGCM calculation based on five axial states is rudimentary, the observed dilatation of spectra is not compensated for by such an enrichment of the PGCM unperturbed state. Correspondingly, the systematic compensation of that dilatation via PGCM-PT(2) corresponds to a genuine action of the perturbation that captures dynamical correlations lying outside the reach of the presently used PGCM ansatz. 
In the end, the PGCM-PT(2) $2^+_1$ excitation energy is independent of $s$ within uncertainties. While reduced compared to PGCM, the $s$ dependence of the PGCM-PT(2)  $4^+_1$ excitation energy is still significant and would probably benefit from being performed on top of a richer PGCM state and/or by going to PGCM-PT(3).

The global picture that emerges for pre-processed Hamiltonians is illustrated for \nucl{Ne}{20} in Fig.~\ref{fig:Ne20_ebranplot}. The MR-IMSRG evolution largely reshuffles the hierarchy of correlations at play. As $s$ grows, one observes that
\begin{enumerate}
\item static correlations captured through the breaking of symmetries at the HFB level as well as by their restoration and the inclusion of collective fluctuations at the PGCM level slightly increase,
\item dynamical correlations brought either on top of HFB via BMBPT(2) or on top of PGCM via PGCM-PT(2) are drastically reduced.
\end{enumerate}
Overall, dynamical correlations go from being highly dominant to being largely subleading. Still, their inclusion on top of PGCM via PGCM-PT(2) remains mandatory, in particular when dealing with low-lying excitation energies.

Eventually, the great benefit of the pre-processing relates to the fact that many-body calculations performed with evolved Hamiltonians become numerically gentler as $s$ increases, i.e. the numerical solution of the PGCM-PT(2) linear system is more precise, corrections beyond PGCM-PT(2) are minimized and the convergence with the model-space size ($e_{\mathrm{max}}$) is probably faster, although this latter point remains to be studied\footnote{See Ref.~\cite{Gebrerufael:2017fk} for an accelerated convergence in so-called in-medium no-core shell model calculations.}. Given that PGCM-PT(2) is numerically more costly than the MR-IMSRG(2) step (see App.~\ref{complexity}), the optimal combination of both methods is of great interest. Of course, this optimal point must be such that the error due to the breaking of unitarity through the MR-IMSRG(2) pre-processing is not larger than the error associated with PGCM-PT(2) results.

\section{Conclusions}
\label{conclusions}

This work, the third paper of the series on PGCM-PT, presented the first realistic results for the novel multi-reference perturbation theory built on top of an unperturbed state generated through the projected generator coordinate method. While the unperturbed state captures crucial static correlations via the breaking and restoration of symmetries along with collective fluctuations, the perturbative expansion brings in complementary dynamical correlations in a consistent fashion within a symmetry conserving scheme. Furthermore, being a {\it state-specific} multi-reference many-body perturbation theory, PGCM-PT accesses ground and low-lying excited states on an equal footing.

First, the novel many-body formalism was shown to be both versatile and accurate by benchmarking proof-of-principle results for the doubly closed-shell \nucl{O}{16}, singly open-shell  \nucl{O}{18} and doubly open-shell  \nucl{Ne}{20} nuclei in a small ($e_{\mathrm{max}}=4$) harmonic oscillator model space against full configuration interaction results. Binding energies obtained at second order, i.e. through PGCM-PT(2), were shown to be typically $0.5-1.5\%$ away from FCI results.

The second focus of the present paper was to demonstrate the benefit of combining low-order PGCM-PT with a pre-processing of the Hamiltonian via multi-reference in-medium similarity renormalization group transformations. The rather low cost of MR-IMSRG(2) calculations makes it possible to efficiently capture the bulk of dynamical correlations in large model spaces (cf.~Refs.~\cite{Gebrerufael:2017fk,Yao:2018qjv,PhysRevLett.124.232501}). Based on such a pre-processed Hamiltonian, PGCM-PT(2) can bring in crucial static correlations {\it and} any remaining dynamical correlations while working in a smaller model space. The present work showed that, after the MR-IMSRG(2) pre-processing, dynamical correlations included on top of the PGCM via PGCM-PT(2) are indeed essential for a satisfactory description of low-lying spectra.

Eventually, it emerges from the present work that a versatile and accurate description of complex mid- and heavy-mass nuclei will probably rely on the combination of three levers whose complementarity needs to be further studied and optimized:
\begin{enumerate}
\item the pre-processing of the Hamiltonian via, e.g., MR-IMSRG to efficiently capture the bulk of dynamical correlations,
\item the use of a, e.g., PGCM unperturbed state capturing collective static correlations via a low-dimensional diagonalization problem that is thus scalable to heavy nuclei\footnote{This can be particularly useful for implementations on GPUs or other accelerators with limited memory.},
\item the low-order truncation of a systematic expansion on top of the multi-reference unperturbed state via, e.g., PGCM-PT to bring in remaining dynamical correlations.
\end{enumerate}
Each of the three steps comes with its own flexibility that can be exploited in order to optimize their combination\footnote{It is worth mentioning that the combination of the three steps is always consistent, i.e. there is no double counting given that each step automatically adapts to the other two.}. First, the pre-processing is a function of a flow parameter $s$ that must be optimized to resum the bulk of dynamical correlations without inducing a large breaking of unitarity\footnote{In this context, the truncation order of the MR-IMSRG(n) procedure plays a critical role. MR-IMSRG(3), for instance, would allow us to reduce any violation of the unitarity, but it comes with a significantly higher numerical cost.}. Second, the PGCM depends on a choice of suitable collective coordinates that must be rich enough to capture all non-perturbative static correlations at play, only leaving weak perturbative corrections to the subsequent PGCM-PT step, while maintaining a low-enough dimensionality to retain its advantage over large-scale diagonalization methods. For example, while adding the triaxial degree of freedom did not impact the dilated PGCM spectrum of $^{20}$Ne at $s=10,20\,\mathrm{MeV}^{-1}$, the use of HFB states obtained while adding a cranking constraint breaking time-reversal invariance~\cite{PhysRevLett.113.162501,Borrajo:PLB2015} typically compresses the PGCM spectrum as demonstrated in MR-EDF calculations and in recent ab initio studies~\cite{PhysRevLett.124.232501,Hergert:2020FP}. Obtaining such a compression at the PGCM level is expected to correlate with a further suppression of dynamical correlations on top of the PGCM step. Still, if needed, the PGCM-PT can in principle be implemented at various perturbative orders $n$. In practice, however, going beyond PGCM-PT(2) shall probably be avoided due to the prohibitive numerical scaling. 

While the present work has laid the foundations of such an optimal scheme, future studies will allow us to better understand the way many-body correlations can be most efficiently captured in complex heavy nuclei within an ab initio setting. For example, describing nuclei displaying strong shape coexistence via ab initio many-body calculations constitute an interesting milestone to achieve in the years to come.

\begin{acknowledgements}
The authors thank M. Saunders and T. Choi for sharing their iterative MINRES-QLP solver and providing useful insights, as well as A. Tichai for interesting discussions. This project has received funding from the European Union’s Horizon 2020 research and innovation programme under the Marie Skłodowska-Curie grant agreement No. 839847. T. R. R. acknowledges the support of the the Spanish MICINN under PGC2018-094583-B-I00. R. R. was supported by the DFG Sonderforschungsbereich SFB 1245 (Project ID 279384907) and the BMBF Verbundprojekt 05P2021 (ErUM-FSP T07, Contract No. 05P21RDFNB). H. H. was supported by the U.S. Department of Energy, Office of Science, Office of Nuclear Physics under Grant No. de-sc0017887. J.M.Y. was supported by the Fundamental Research Funds for the Central Universities, Sun Yat-sen University.
The MR-IMSRG code uses optimizations done in collaboration with the U.S. Department of Energy, Office of Science, Office of Advanced Scientific Computing Research, Scientific Discovery through Advanced Computing (SciDAC) Program FASTMath and RAPIDS2 Institutes. Calculations were performed by using HPC resources from GENCI-TGCC (Contract No. A009057392), CETA-Ciemat (FI-2021-2-0013), the Lichtenberg high performance computer of the Technische Universit\"at Darmstadt and the Institute for Cyber-Enabled Research (ICER) at Michigan State University.
\end{acknowledgements}

\begin{appendices}

\section{Anti-symmetry reduction}

From a technical viewpoint, and as extensively explained in Paper I, PGCM-PT(2) calculations rely on solving a large-scale linear problem of the form
\begin{align}
{\bold M} {\bold a} & =  -{\bold h_1} \, . \label{eq:lin_num}
\end{align}

The linear problem relates to excitations $I$ of several non-orthogonal Hartree-Fock-Bogoliubov vacua, each of which is defined by a set of quasi-particle creation operators, i.e. a rank-$n$ excitation is defined through a set of $n$ quasi-particle labels \(I\sim (k_{i_1}, \cdots, k_{i_n})\). Correspondingly, the problem is initially expressed in terms of unrestricted sets of quasi-particle indices. Still,  anti-commutation relations of quasi-particle creation operators imply that \( \mathbf M,\  \mathbf a\) and \( {\bold h_1}\) are anti-symmetric with respect to the permutations of quasi-particle indices. This can be exploited to reduce the effective dimensionality of the linear system.

Given a rank-$n$ excitation \(I\sim (k_{i_1}, \cdots, k_{i_n})\) on a given Bogoliubov state, the set \(\mathcal I \equiv \{\tau( I)\}_{\tau\in \mathcal S_n}\) of \(|\mathcal I| \equiv n!\) permutations of the quasi-particle indices of \(I\) needs to be considered. For a pair \((I,J)\) of excitations and two permutations \((\tau,\tau')\) applicable to \(I\) and \(J\), the antisymmetry properties are given by
\begin{subequations}
\label{antisymmatrices}
    \begin{align}
          M _{pIqJ} &= \epsilon(\tau) \epsilon(\tau')   M_{p\tau(I) q\tau'(J)} \, ,\\
          a^{J}(q) &=\epsilon(\tau)   a^{\tau(J)}(q) \, ,\\
          h_1^{I}(p) &= \epsilon(\tau')   h_1^{\tau'(I)}(p) \, ,
    \end{align}
\end{subequations}
where $\epsilon(\tau)$ denotes the signature of the permutation $\tau$. First, these antisymmetry properties trivially imply that excitations with repeated quasi-particle indices can be excluded from the basis. Second, the set of excitations \(\mathcal I\) corresponding to one another via a change of the quasi-particle ordering can all be tracked through the one representative \(\bar I\) of \(\mathcal I\) characterized by a strictly increasing ordering of the quasi-particle indices \(k_1 < \cdots < k_n\). Writing Eq. \eqref{eq:lin_num}  for such an external ordered excitation \(\bar I\)
\begin{align}
        \sum_{q} \sum_{\mathcal J} \sum_{J\in \mathcal J}
         M _{p\bar I qJ}  a^{J}(q) &= - h_1^{\bar I}(p) \, ,
\end{align}
the internal sum is split such that, with the help of Eq.~\eqref{antisymmatrices}, $|\mathcal J|$ equivalent terms are generated that eventually yield the reduced form 
\begin{align}
        \sum_{q} \sum_{\mathcal J} |\mathcal J|
         M _{p\bar I\ q \bar J}  a^{\bar J}(q) &= - h_1^{\bar I}(p) \, .
\end{align}
In order to maintain the Hermiticity of the reduced matrix one further left-multiplies the equation by \(\sqrt|\mathcal I|\) such that the final form
        \begin{align}
        \sum_{q} \sum_{\mathcal J} \sqrt{|\mathcal J|}\sqrt{|\mathcal I|}
         M _{p\bar I q\bar J} \sqrt{|\mathcal J|} a^{\bar J}(q) &= -\sqrt{|\mathcal I|} h_1^{\bar I}(p) \, ,
\end{align}
naturally leads to a trivial redefinition of the reduced matrix and vectors through the inclusion of the combinatorial factors. In the following, the above reduction process is assumed such that the effective working basis only includes excitations characterized by quasi-particle indices in a strictly increasing order. For example, exploiting the anti-symmetry for the dominant double (i.e. 4 quasi-particle) excitations reduces the number of associated matrix elements by a factor of \(24^2\).

\section{Solution of the linear problem}
\label{solvingLS}

Finding the numerical solution of Eq. \eqref{eq:lin_num} is delicate due to the non-orthogonality of the many-body states used to represent it. Thus, a careful handling of zeros in the norm eigenvalues is typically necessary to avoid instabilities while solving the equation. In the following, techniques of increasing sophistication are progressively introduced in order to eventually motivate the use of the iterative MINRES-QLP algorithm.

\subsection{Exact SVD-based solution}

The pedestrian way to solve the linear system can be summarized in three steps: (i) diagonalize the norm matrix to transform the equation into an orthonormal basis, (ii) diagonalize the Hamiltonian matrix in that basis and (iii) finally invert the problem. This strategy is essentially the same as the one used in PGCM to solve the HWG equation (see App.~A of Paper II).

The norm matrix (see Paper I) is first decomposed by projecting on the range of \(\mathbf N\) via a singular-value decomposition (SVD)
\begin{equation}
    \mathbf N = \mathbf X\mathbf I \mathbf X^\dag \, ,
\end{equation}
where $\mathbf X$ is unitary. Matrix $I$ gathers the singular values whose smallest representatives can eventually be discarded. Correspondingly, $\mathbf M$, $\mathbf a$ and $\mathbf h_1$ are transformed into the resulting orthogonal basis
\begin{subequations}
\begin{align}
  \mathbf  { \tilde{M}} &\equiv \mathbf X \mathbf M \mathbf X^\dag \, ,\\
   \mathbf {\tilde{ a}} &\equiv \mathbf X^\dag \mathbf a \, ,\\
    \mathbf {\tilde{h}}_1 &\equiv \mathbf X^\dag \mathbf  h_1 \, ,
\end{align}
\end{subequations}
such that the linear problem equivalently reads
\begin{equation}
    \mathbf  { \tilde{M}}  \mathbf {\tilde{ a}}  = - \mathbf {\tilde{h}}_1 \, .
\end{equation}
The solution of this system is then found by diagonalizing \(\mathbf {\tilde M}\)
\begin{equation}
    \mathbf \Delta = \mathbf Y^\dag \mathbf {\tilde M} \mathbf Y \, ,
\end{equation}
such that, similarly to canonical MBPT, the system is inverted in the basis where \( \mathbf {\tilde M} \) is diagonal to obtain the second-order energy in the form
\begin{equation}\label{eq:E2_pedestrian}
    E^{(2)} = - \mathbf { h_1}^\dag \mathbf X\mathbf Y \mathbf\Delta^{-1}\mathbf Y^\dag\mathbf X^\dag \mathbf { h_1} \, .
\end{equation}
In principle, the projection on the range of \(\mathbf N\) is not necessary to solve the system. However, in numerical applications, the coupling between spurious eigenvalues of \(\mathbf N\) and large eigenvalues of \(\mathbf M\) can arise and the explicit removal of the redundancies is often necessary. Eventually, full diagonalization is anyway not feasible for the large matrices encountered in realistic PGCM-PT(2) calculations (contrary to the PGCM step where the typical dimensions are sufficiently small) such that other methods need to be designed to solve the problem.

\begin{figure}
    \includegraphics[width=.5\textwidth]{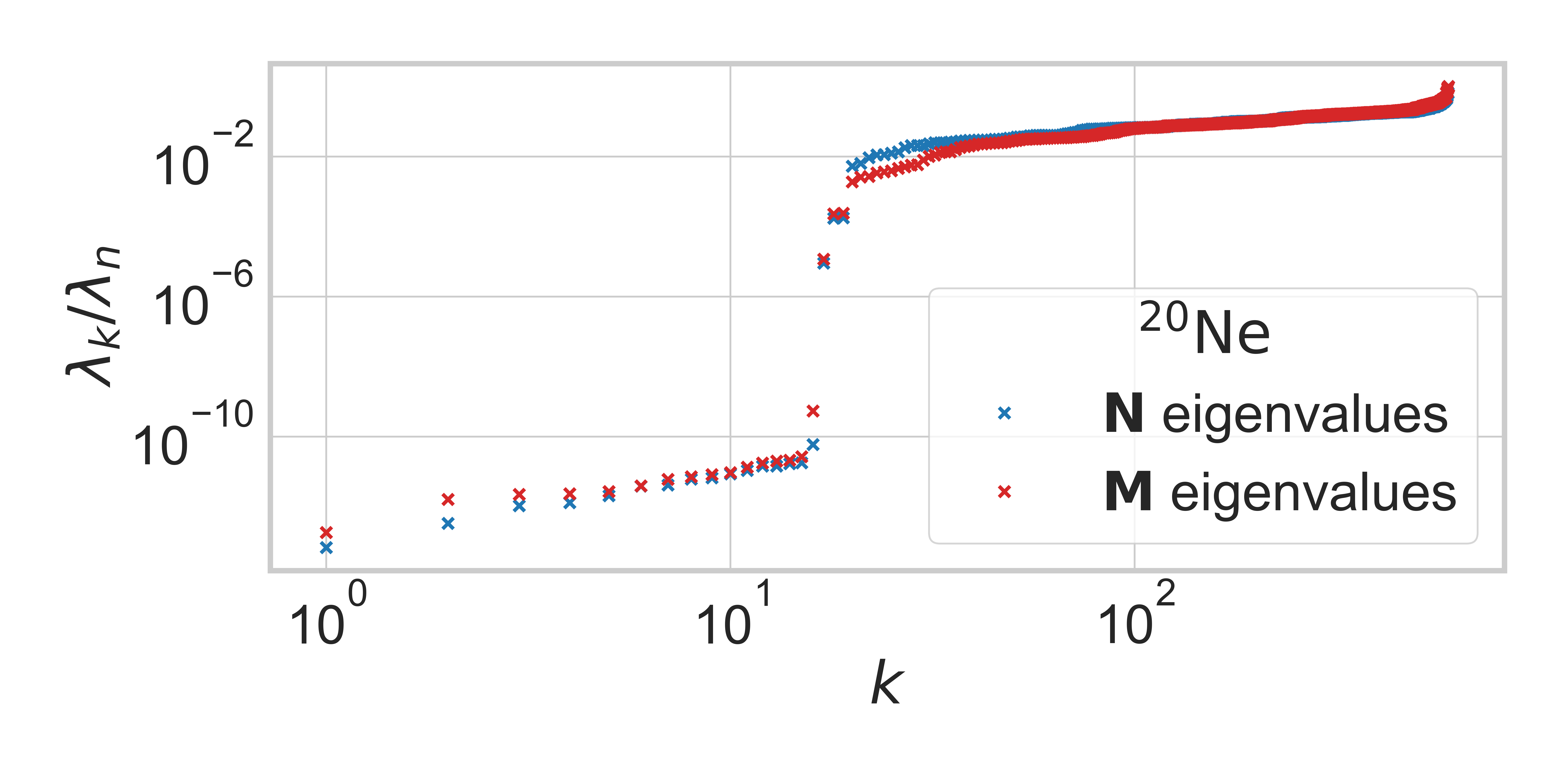}
    \caption{(Color online) Distribution of eigenvalues of \(\mathbf N\) and \(\mathbf M\) matrices for \nucl{Ne}{20}. The calculation is performed with a two-body \(\chi\)EFT Hamiltonian, $\lambda_{\mathrm{srg}}=1.88\,\mathrm{fm}^{-1}$, $\hbar \omega=20\,\mathrm{MeV}$ and $e_{\mathrm{max}}=2$.}
    \label{fig:spectra_MN}
\end{figure}

As an example, the distribution of the eigenvalues of \(\mathbf N\) and \(\mathbf M\) obtained from a PHFB-PT(2) calculation of \nucl{Ne}{20} in a small model-space is displayed in Fig.~\ref{fig:spectra_MN}. Two observations can be made
\begin{itemize}
    \item The eigenvalue distributions of both matrices are very close up to a scaling factor. In particular, as expected, their (numerical) kernels have the same dimension.
    \item The kernel's dimension is small compared to the matrices' dimension, and all eigenvalues outside the kernel have the same magnitude. This prevents us from using truncated SVD approaches in larger model spaces.
\end{itemize}
Although the GCM mixing enlarges the kernel of the PGCM-PT(2) matrices compared to PHFB-PT(2) due to the partial linear dependencies of the added HFB states, a large number of independent configurations is still present in that case too.

\subsection{Pivoted factorizations}

Factorization algorithms can be applied in order to remove spurious eigenvalues without paying the price of fully diagonalizing the norm and Hamiltonian matrices.
Typical examples are pivoted QR~\cite{Businger1971} and QLP~\cite{QLP} factorizations, which are briefly discussed in the following.

\subsubsection{Pivoted QR}

An arbitrary \(n\times m\) complex matrix \(A\) can be decomposed according to
\begin{equation}
    A D = Q R \, ,
\end{equation}
where \(D\) is obtained via a permutation of the columns of \(A\), \(Q\) is a unitary matrix, and \(R\) is an upper triangular matrix. The permutation \(D\) is used to sort the diagonal entries of \(R\) in decreasing order of magnitude. In this way, the kernel of \(A\) corresponds to the last columns of \(R\).

\subsubsection{Pivoted QLP}

Matrix A can be decomposed further by performing two successive pivoted QR decompositions, yielding the form
\begin{equation}
    D' A D = QLP
\end{equation}
where \(D, D'\) are permutation matrices, \(Q, \ P\) are unitary matrices and \(L\) is a lower triangular matrix. In particular, \(L\) has the block-diagonal form
\begin{equation}
      L = \begin{pmatrix}
        \tilde L & 0\\
        0 & 0
      \end{pmatrix}
    ,
\end{equation}
such that \(Q\) and \(P\) naturally block factorize \(A\) into a full-rank part and its null-space.

\subsubsection{Illustration of non-iterative solvers}

\begin{figure}
    \includegraphics[width=.5\textwidth]{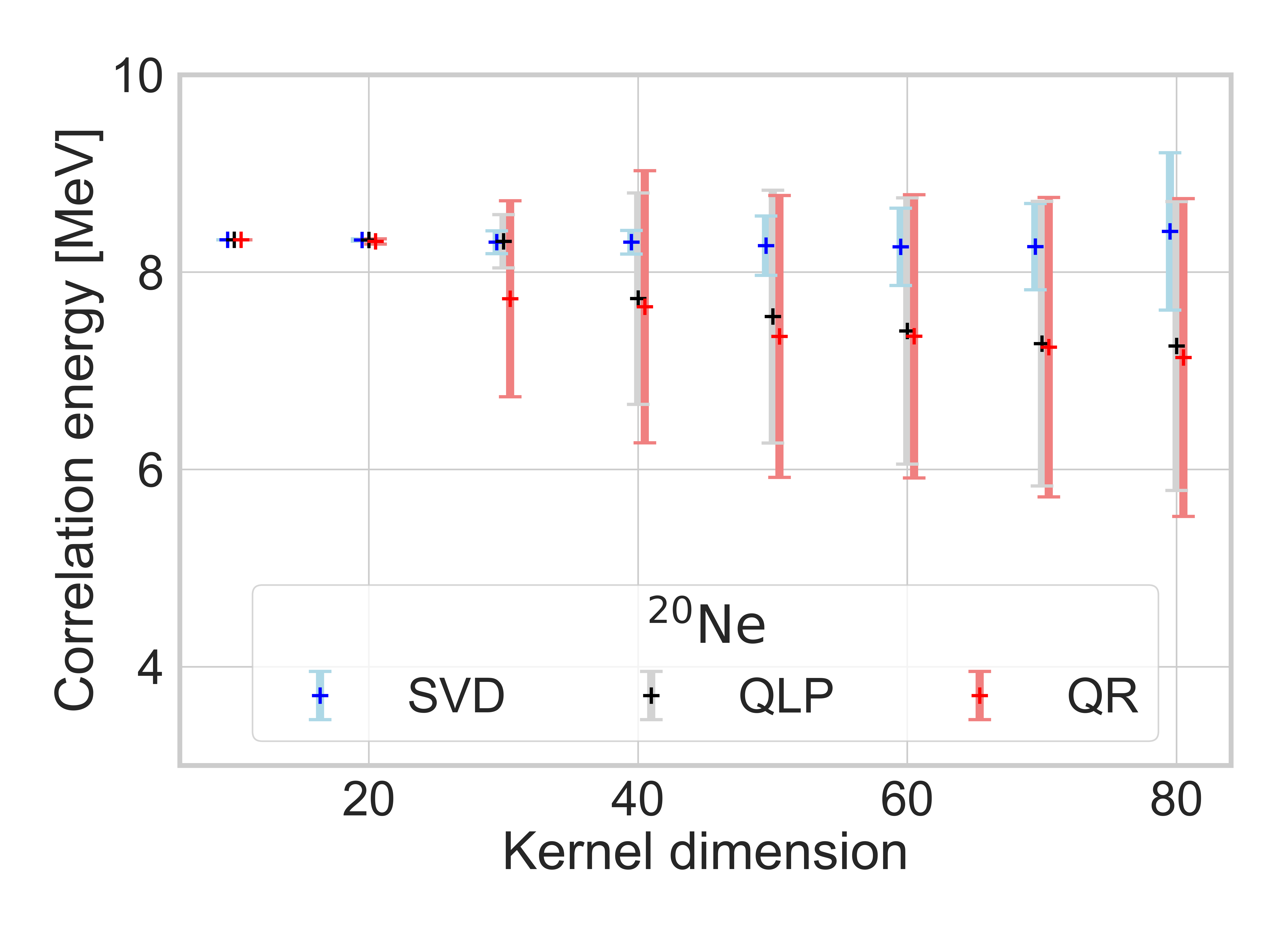}
    \caption{(Color online) PHFB-PT correlation energy of \nucl{Ne}{20} obtained for SVD, QR and QLP decompositions as function of the size of the excluded kernel in the decomposition. The calculation is performed with a two-body \(\chi\)EFT Hamiltonian, $\lambda_{\mathrm{srg}}=1.88\,\mathrm{fm}^{-1}$, $\hbar \omega=20\,\mathrm{MeV}$ and $e_{\mathrm{max}}=2$.}
    \label{fig:exact_decomposition}
\end{figure}

In our case, pivoted QR/QLP factorizations can either be used directly on \(\mathbf M\) to solve Eq. \eqref{eq:lin_num} or on the norm matrix in order to remove redundancies in the basis. In both cases, the symmetry of the matrices guarantees that the range and the kernel of both matrices are in direct sum. QLP factorization can thus be seen as a way to re-express the original problem in the range of \(\mathbf M\) or \(\mathbf N\).
In practical applications, some tolerance must be used (as with SVD) to discard numerically small eigenvalues and disentangle the numerical kernel from the numerical range. The QLP factorization, although twice more expensive than the single QR decomposition, is found to be more stable and to better discard spurious eigenvalues.

Figure~\ref{fig:exact_decomposition} shows a comparison of SVD, QR and QLP decompositions in a PHFB-PT(2) calculation. Errors estimated via \(\delta E\equiv \|\mathbf M \mathbf a + \mathbf h_1\|\|\mathbf a\|\) (see Sec.~\ref{erroreval}) are also represented on the figure. The three methods are in very good agreement with vanishing errors when nearly all the space is kept in the calculation. However, discrepancies arise when the truncation is performed according to the magnitude of the diagonal elements of the decomposition. While the SVD is by far the most reliable method, the QLP decomposition significantly improves the correlation energy with respect to the simpler QR decomposition and reduces the corresponding error for a fixed kernel dimension. Eventually, the reduced cost of QLP/QR decompositions compared to SVD, especially in their sparse version, make them well suited to large-scale calculations.

\subsection{Iterative solvers}

The QLP decomposition introduced above is still not applicable to very large matrices due to the runtime and memory requirements. An alternative solution is to use an iterative method, preferably exploiting the symmetry of the input matrix. Among various available solvers, the MINRES algorithm~\cite{MINRES} finds the minimum-residual solution to \(||\mathbf M \mathbf a + \mathbf h_1||\) via QR factorizations in the Krylov space of \(\mathbf M\). In the case of ill-conditioned problems, QR factorizations are replaced by QLP factorizations, and the corresponding algorithm is referred to as MINRES-QLP~\cite{MINRES-QLP}.

The benefit of iterative solvers compared to exact decompositions is that, in the former, QR and QLP factorizations are performed in a Krylov subspace of the matrix. At iteration \(k\), the problem is of dimension \(k\times k\), where \(k\) is usually much smaller than the original matrix dimensions. This results in both runtime and memory savings, at the cost of solving the system only approximately.

\subsubsection{Preconditioning of the linear system}
\label{subsec:precond}

The number of iterations required by the solvers strongly depends on the eigenvalue distribution of the linear system under consideration. Typically, systems where the eigenvalues are clustered will have a faster convergence than systems with a spread-out spectrum. The spread of the eigenvalues can be altered with preconditioning techniques that amounts to finding equivalent systems with different (generally much smaller) condition numbers.

In this subsection, the compatible symmetric system
\begin{equation}
    \mathbf M \mathbf x = -\mathbf h_1 \, ,
\end{equation}
is considered. Let \(\mathbf A = \mathbf C \mathbf C^T\) be a positive definite matrix. The solution of the initial system can be deduced from the solution of the preconditioned system
\begin{equation}
    \mathbf C^{-1} \mathbf M \mathbf C^{-T} \mathbf y = - \mathbf C^{-1} \mathbf h_1 \, .
\end{equation}
 Whereas various techniques are available to build an appropriate matrix \(\mathbf A\), designing efficient preconditioners is still an active field of research \cite{WATHEN}. There is no perfect preconditioner, and finding the trade-off between effectiveness and computational cost heavily relies on heuristics. Furthermore, for systems only known up to a given precision, preconditioners can artificially magnify eigenvalues that are numerically close to zero. Thus, a slower convergence with the possibility to stop the iterations before the appearance of spurious divergences might be preferable. Eventually, solving several equivalent systems simultaneously can make it easier to identify problematic features, and discrepancies between different solutions can be used as uncertainty estimates in the resolution.

\paragraph{Matrix scaling.}

Matrix scaling is a type of preconditioning where the preconditioner is a diagonal matrix \(\mathbf A \equiv \mathbf D\) such that the equivalent system reads
\begin{equation}
    \mathbf D^{-\inv2} \mathbf M \mathbf D^{-\inv2} \mathbf y = - \mathbf D^{-\inv2} \mathbf h_1 \, .
\end{equation}
If the scaled matrix
\begin{equation}
    \mathbf {\tilde M} \equiv \mathbf D\mathbf M\mathbf D
\end{equation}
is better conditioned than \(\mathbf M\), then the solution to the initial system can be found in fewer iterations. For a diagonally dominant matrix \(\mathbf M\), scaling the matrix with its own diagonal elements will reduce its condition number. The binormalization method detailed in Ref.~\cite{BINORM} amounts to scaling all rows and columns to unit norm, which can yield significantly better results at low cost. A stochastic matrix-free variant \cite{Bradley2010} allows one to efficiently apply this method for abstract linear operators that are not necessarily defined explicitly by a matrix-vector product. Below, the stochastic binormalization preconditioner is denoted as SBIN.

\paragraph{Incomplete Cholesky decomposition.}

For a sparse positive definite matrix \(\mathbf N\), an approximate Cholesky factorization preserving the sparsity pattern of the original matrix can be computed as
\begin{equation}
    \mathbf N \sim \mathbf L \mathbf L^T,
\end{equation}
with \(\mathbf L\) a (sparse) lower triangular matrix. A variant of Cholesky factorization applicable to positive indefinite matrices can be applied directly on the norm matrix \(\mathbf N\). Since \(\mathbf N\) and \(\mathbf M\) have similar eigenvalue spread, eigenvalues of the system preconditioned with \(\mathbf L\mathbf L^T\) will be much more clustered than those of the original system, hence separating the useful directions of the problem from the rest of the Hilbert space. Below, the incomplete Cholesky preconditioner is denoted as IC0.

\paragraph{Norm preconditioning.}

In some cases, spurious eigenvalues in the linear system can couple to physical modes and prevent any convergence of the iterative solvers. In this case, clustering the eigenvalues via preconditioning techniques is counterproductive as spurious modes are given an equivalent amplitude to physical ones. When this happens, it is preferable to amplify the separation of scale between numerically small and large eigenvalues. Instead of manually removing redundancies in the norm matrix, there exists a simple way to reach the image of \(\mathbf N\) without resorting to a decomposition: Instead of solving 
\begin{equation}
    \mathbf M \mathbf a = - \mathbf h_1 \, ,
\end{equation}
one directly solves for \(\mathbf N^{-1} \mathbf a\) inside the range of \(\mathbf N\) via
\begin{equation}
    \mathbf N \mathbf M \mathbf N \left(\mathbf N^{-1} \mathbf a\right) = - \mathbf N \mathbf h_1 \, .
\end{equation}
Even if \(\mathbf N\) is singular, the fact that \(\mathbf h_1 \text{ and } \mathbf a\) live in the range of \(\mathbf N\) by construction ensures that \(\mathbf N^{-1} \mathbf a\) is well-defined.
The procedure ensures that small numerical eigenvalues of \(\mathbf N\), originating from collinear many-body basis vectors, are tamed down in \( \mathbf N \mathbf M \mathbf N \). Furthermore, numerical errors in \(\mathbf h_1\) are suppressed as well. Of course, in exact arithmetic, the two systems are equivalent. This method corresponds in fact to preconditioning the system with \(\mathbf N^{-2}\). As mentioned, this slows down the convergence of the iterative procedure and must be kept for cases where the direct approach or the complex shift method (see below) do not provide accurate solutions. Below, the norm precondition is denoted as \( \mathbf N \mathbf M \mathbf N \).

\subsubsection{Error evaluation}
\label{erroreval}

Iterative methods may require a large number of iterations or even diverge due to numerical errors. In this subsection, a conservative bound to estimate the error on the computed second-order energy is developed.

Given an approximate solution of the system
\begin{equation}
    \mathbf M \mathbf a = -\mathbf h_1 +\mathbf b \, ,
\end{equation}
the second-order energy evaluated with Hylleraas' functional reads
\begin{align}
    E^{(2)} &= 
    \mathbf a^\dag \mathbf M \mathbf a + \mathbf h_1^\dag \mathbf a + \mathbf a^\dag \mathbf h_1 \nonumber \\
            &= 
    \mathbf a^\dag \mathbf b + \mathbf h_1^\dag \mathbf a \, .
\end{align}
The difference between this expression and the directly evaluated second-order energy reads 
\begin{equation}
    \delta E^{(2)} = \mathbf a^\dag \left(\mathbf M\mathbf a + \mathbf h_1\right) \, .
\end{equation}
Thus, a conservative error estimate on the second-order energy is given by
\begin{equation}
    |\delta E|^{(2)} \equiv \| \mathbf a\| \|\mathbf M \mathbf a + \mathbf h_1\| \, . \label{estimatederror}
\end{equation}
The quantity $|\delta E|^{(2)}$ vanishes for an exact solution and grows whenever \(\|\mathbf a \|\) becomes too large, which generally occurs if \(\mathbf M\) is badly conditioned. 
When the norm-preconditioning is used, the error estimate is  obtained as
\begin{equation}
    |\delta E|^{(2)} \equiv \| \mathbf N^{-1} \mathbf a\| \|\mathbf N \mathbf M \mathbf N \left(\mathbf N^{-1}\mathbf a\right) + \mathbf N \mathbf h_1\| \, .
\end{equation}

\subsubsection{Stopping condition for the iterative solver}

MINRES-QLP already implements by default its own stopping criterion based on the relative norm of the residuals
\begin{equation}
    r\equiv\frac{\|\mathbf M \mathbf a + \mathbf h_1\|}{\|\mathbf a\|}.
\end{equation}
In the present case, elements of \(\mathbf M\) and \(\mathbf h_1\) are obtained after several computational steps such that round-off and discretization errors will alter the quality of the input matrices. Furthermore, a threshold on the magnitude of the matrix elements of \(\mathbf M\) is employed to enforce the sparsity of the matrix. As such, iterations should be stopped when the residual errors are of the same order as the precision of the input matrix elements.

\subsubsection{Illustration of iterative solvers}

\begin{figure}
    \centering
    \includegraphics[width=.5\textwidth]{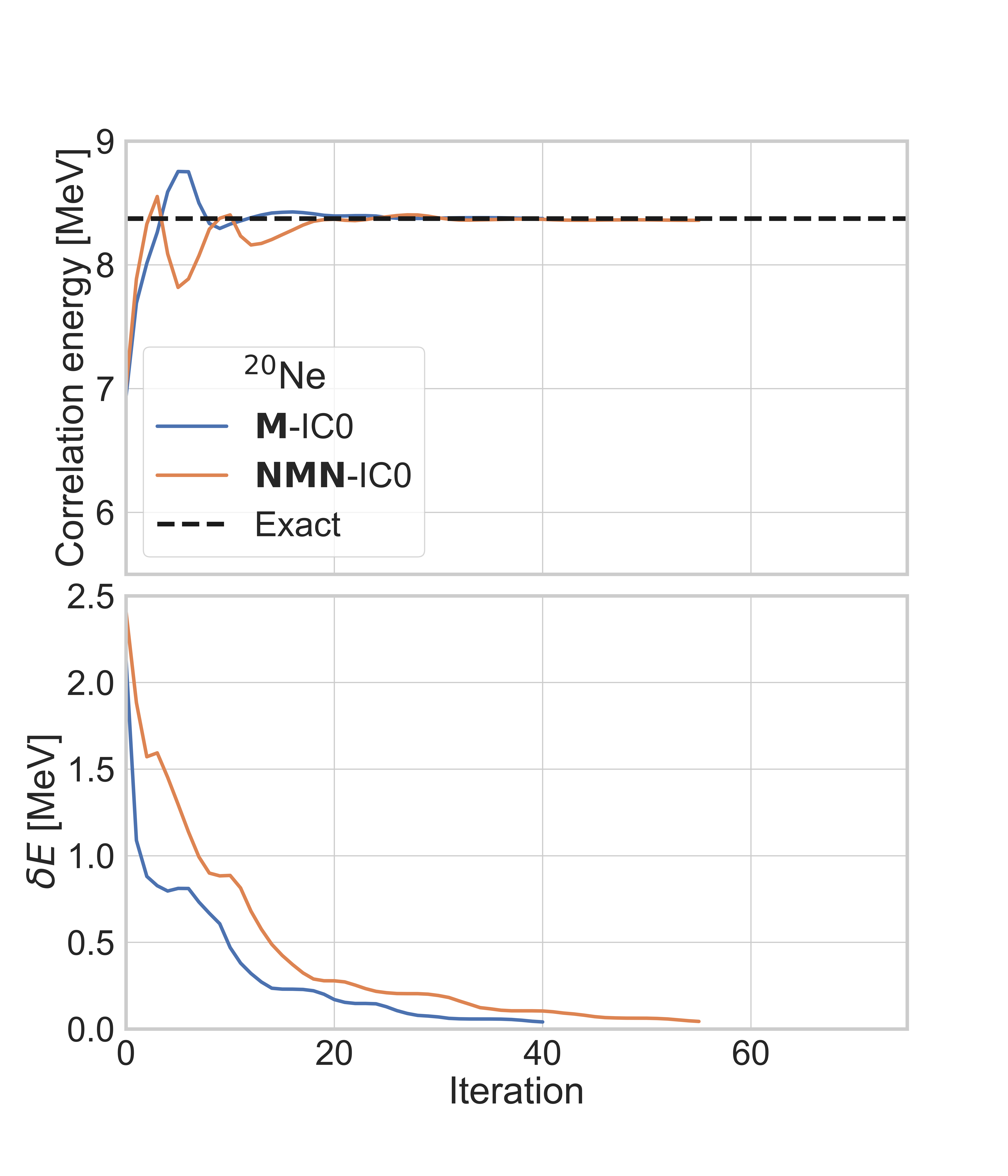}
    \caption{(Color online) Correlation energy (top) and corresponding error (bottom) at each MINRES-QLP iteration for the ground state of \nucl{Ne}{20}. Results obtained with  combination of IC0 and norm preconditionings are compared to the exact solution obtained via SVD. Calculations are performed with a two-body \(\chi\)EFT Hamiltonian~\cite{Huther_2020,Entem:2017gor}, $\lambda_{\mathrm{srg}}=1.88\,\mathrm{fm}^{-1}$, $\hbar \omega=20\,\mathrm{MeV}$ and $e_{\mathrm{max}}=2$.}
    \label{fig:Ne20_minres-qlp}
\end{figure}

\begin{figure}
    \centering
    \includegraphics[width=.5\textwidth]{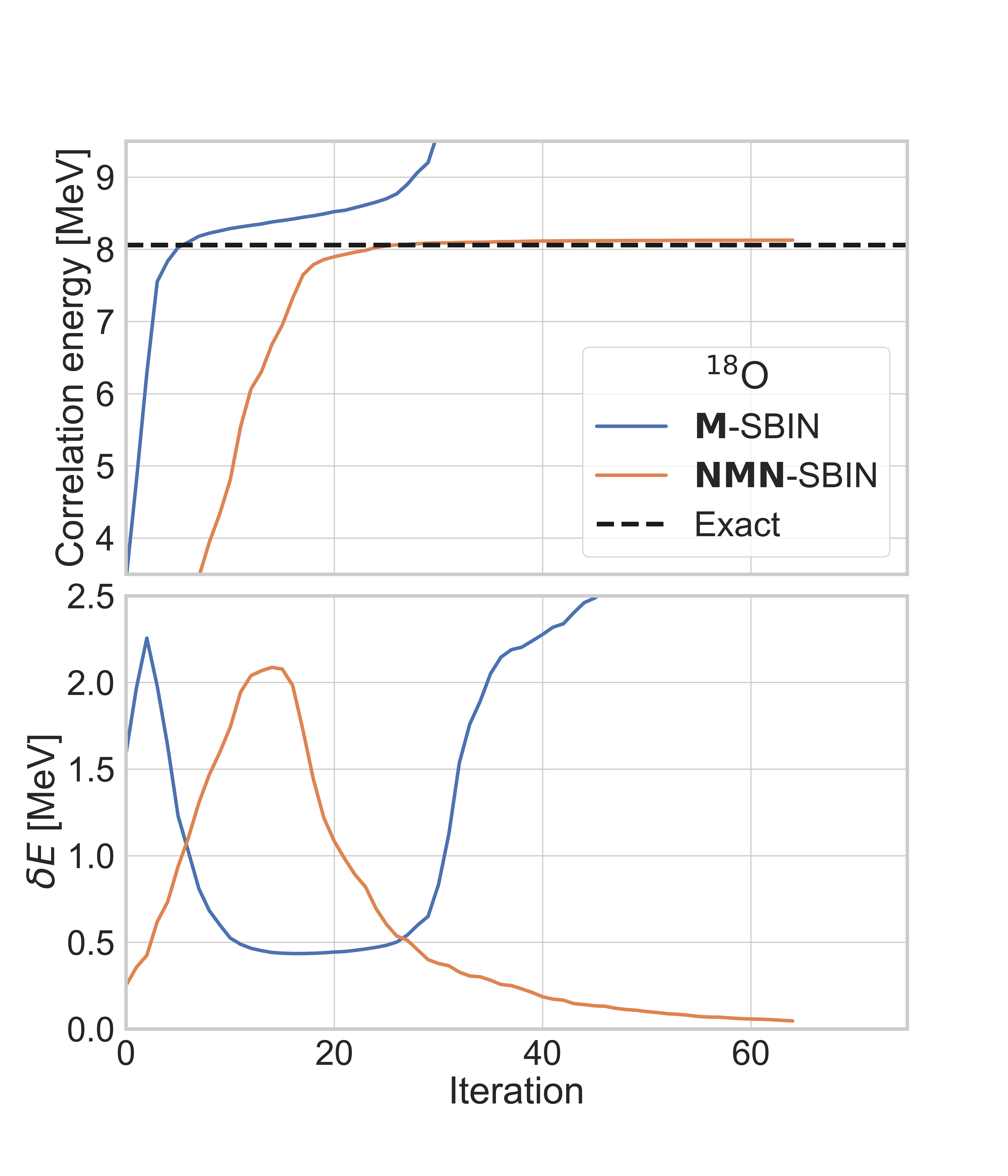}
    \caption{(Color online) Correlation energy (top) and corresponding error (bottom) at each MINRES-QLP iteration for the ground state of \nucl{O}{18}. Results obtained with a combination of SBIN and norm preconditionings are compared to the exact solution obtained via SVD. Calculations are performed with a two-body \(\chi\)EFT Hamiltonian~\cite{Huther_2020,Entem:2017gor}, $\lambda_{\mathrm{srg}}=1.88\,\mathrm{fm}^{-1}$, $\hbar \omega=20\,\mathrm{MeV}$ and $e_{\mathrm{max}}=2$.}
    \label{fig:O18_minres-qlp}
\end{figure}

In order to illustrate the use of iterative solvers and preconditioning techniques, results obtained in \nucl{Ne}{20} with  $e_{\mathrm{max}}=2$ are shown in Fig.~\ref{fig:Ne20_minres-qlp}. One observes that the IC0 preconditioning significantly reduces the number of iterations needed to reach the converged value. Contrarily, the norm preconditioning tends to spread the eigenvalues of the system and therefore slows down the convergence. For a well-behaved system, applying the IC0 preconditioning to the original matrix is therefore the method of choice. In contrast, whenever spurious eigenvalues prevent the convergence of the iterative process, the IC0 preconditioning amplifies the problem. Such a case is shown Fig. \ref{fig:O18_minres-qlp} for the ground state of \nucl{O}{18}. Here, applying a combination of SBIN and norm preconditioning is necessary to converge the system to the SVD solution.

\section{Complex-shift method\label{sec:shift}}

\subsection{Motivations}

As it appears in Eq. \eqref{eq:E2_pedestrian}, the second-order energy relies on the invertibility of \(\mathbf \Delta\) to generate non-zero energy denominators. However, the eigenvalues of \(\mathbf \Delta\) can vanish, which makes the calculation of the second-order energy unstable or even ill-defined. Multi-reference methods are indeed susceptible to this so-called intruder-state problem~\cite{burton20a,FORSBERG1997}.

One way to regularize these zeros is to introduce a diagonal imaginary shift in the eigenbasis of \(\mathbf M\). The eigenvalues are thus replaced by
\begin{equation}
    \bar{\mathbf \Delta} \equiv \mathbf \Delta + \imath \gamma I \, ,
\end{equation}
or, equivalently, working in the original basis
\begin{equation}
    \bar{\mathbf M} \equiv \mathbf M + \imath \gamma \mathbf N \, ,
\end{equation}
which simply corresponds to adding a complex term to the unperturbed Hamiltonian $H_0$. The imaginary shift
moves zero eigenvalues of \(\mathbf \Delta\) into the complex plane and provides a
robust way to remove intruder-state divergences.

In this context, the second-order energy is eventually evaluated by simply taking the real part of the Hylleraas functional,
\begin{equation}
    E^{(2)} = \Re\left[\mathbf a^\dag \bar {\mathbf M} \mathbf a
    +\mathbf a^\dag \mathbf h_1 + \mathbf h_1^\dag \mathbf a\right].
\end{equation}

\subsection{Implementation in real arithmetic}

Although an extension of MINRES-QLP has been developed to handle complex symmetric matrices \cite{CS-MINRES-QLP,MINRES-IMPL}, it is possible to rewrite the complex PGCM-PT(2) equations as an enlarged real-valued system, for which the original MINRES-QLP algorithm can be applied  directly.
The system of equations 
\begin{equation}
    (\mathbf M + \imath \gamma \mathbf N) (\mathbf a + \imath \mathbf b) = - \mathbf h_1
\end{equation}
is recast into a blockwise 2x2 real symmetric system
\begin{equation}
\begin{pmatrix}
    \mathbf M & -\gamma \mathbf N\\
    -\gamma \mathbf N & -\mathbf M
\end{pmatrix}
\begin{pmatrix}
   \mathbf a\\\mathbf b
\end{pmatrix}
=
\begin{pmatrix}
    -\mathbf h_1\\\mathbf 0
\end{pmatrix} \, .
\end{equation}
Since the matrices are real by default after projection, implementing the imaginary shift via an augmented real system is profitable to make use of the MINRES-QLP real symmetric solver instead of variants designed for complex matrices.

Note that preconditioning techniques such as SBIN and \(\mathbf{N}\)-IC0 are still applicable within the augmented system.

\subsection{Illustration}
\label{sec:shift_ill}

\begin{figure}
    \centering
    \includegraphics[width=.5\textwidth]{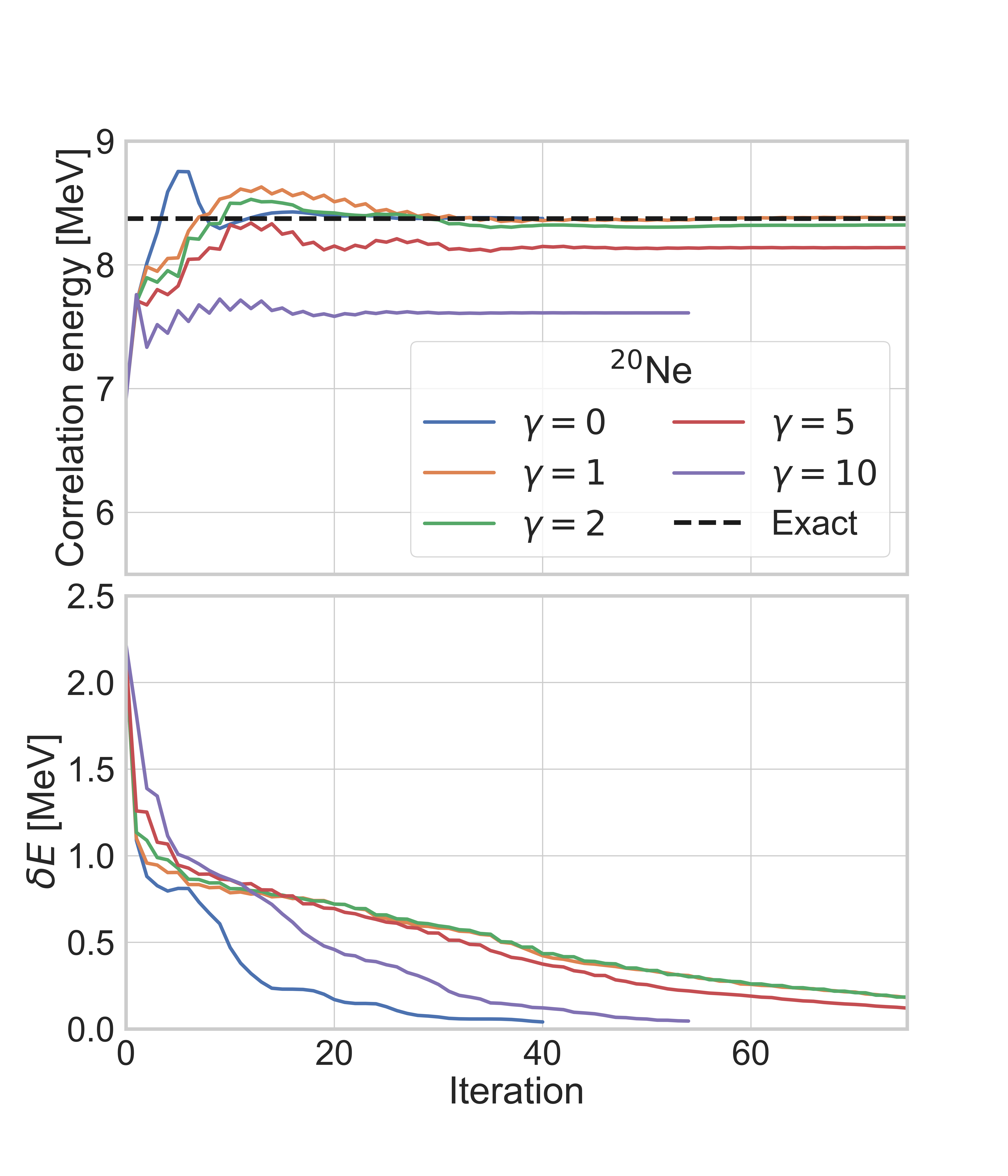}
    \caption{(Color online) Correlation energy (top panel) and corresponding estimated error for different values of the complex shift \(\gamma\) as a function of the number of MINRES-QLP iterations in \nucl{Ne}{20}. Calculations are performed with a two-body \(\chi\)EFT Hamiltonian~\cite{Huther_2020,Entem:2017gor}, $\lambda_{\mathrm{srg}}=1.88\,\mathrm{fm}^{-1}$, $\hbar \omega=20\,\mathrm{MeV}$ and $e_{\mathrm{max}}=2$.}
    \label{fig:Ne20_minres-qlp_cx}
\end{figure}

Proceeding with the \nucl{Ne}{20} test case ($e_{\text{max}}=2$), the effect of the complex shifts on the MINRES-QLP iterations with the IC0 preconditioning is illustrated in Fig.~\ref{fig:Ne20_minres-qlp_cx}. In general, the complex shift tends to lower the correlation energy --- in the limit of an infinite shift, the correlation energy vanishes. Thus, a bias is introduced that must be monitored. Eventually, the larger the complex shift, the faster the iterative procedure will converge (towards a biased result).

\begin{figure}
    \centering
    \includegraphics[width=.5\textwidth]{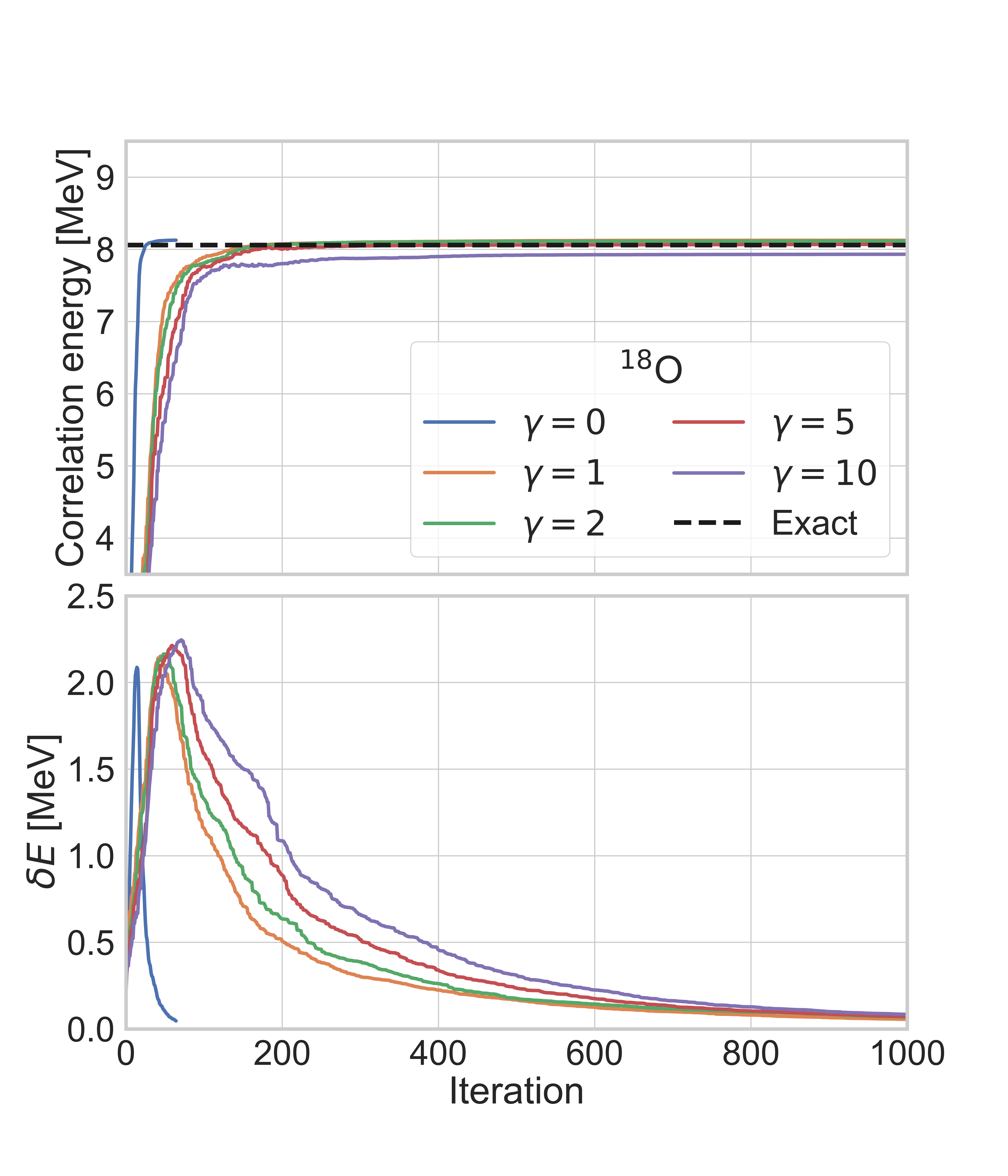}
    \caption{(Color online) Correlation energy (top panel) and corresponding estimated error for different values of the complex shift \(\gamma\) as a function of the number of MINRES-QLP iterations in \nucl{O}{18}. Calculations are performed with a two-body \(\chi\)EFT Hamiltonian~\cite{Huther_2020,Entem:2017gor}, $\lambda_{\mathrm{srg}}=1.88\,\mathrm{fm}^{-1}$, $\hbar \omega=20\,\mathrm{MeV}$ and $e_{\mathrm{max}}=2$.}
    \label{fig:O18_minres-qlp_cx}
\end{figure}

In the case of \nucl{O}{18}, we can combine the norm and SBIN preconditioning (cf.~ Fig.~\ref{fig:O18_minres-qlp}), with the complex shift, as pictured in Fig.~\ref{fig:O18_minres-qlp_cx}. In contrast to \nucl{Ne}{20}, the complex shift with the norm preconditioning decreases the convergence speed in this case. Applying the complex shift without norm preconditioning is also possible, but does only lead to a stabilization of the result before the occurrence of a divergence.

\begin{figure}
    \centering
    \includegraphics[width=.5\textwidth]{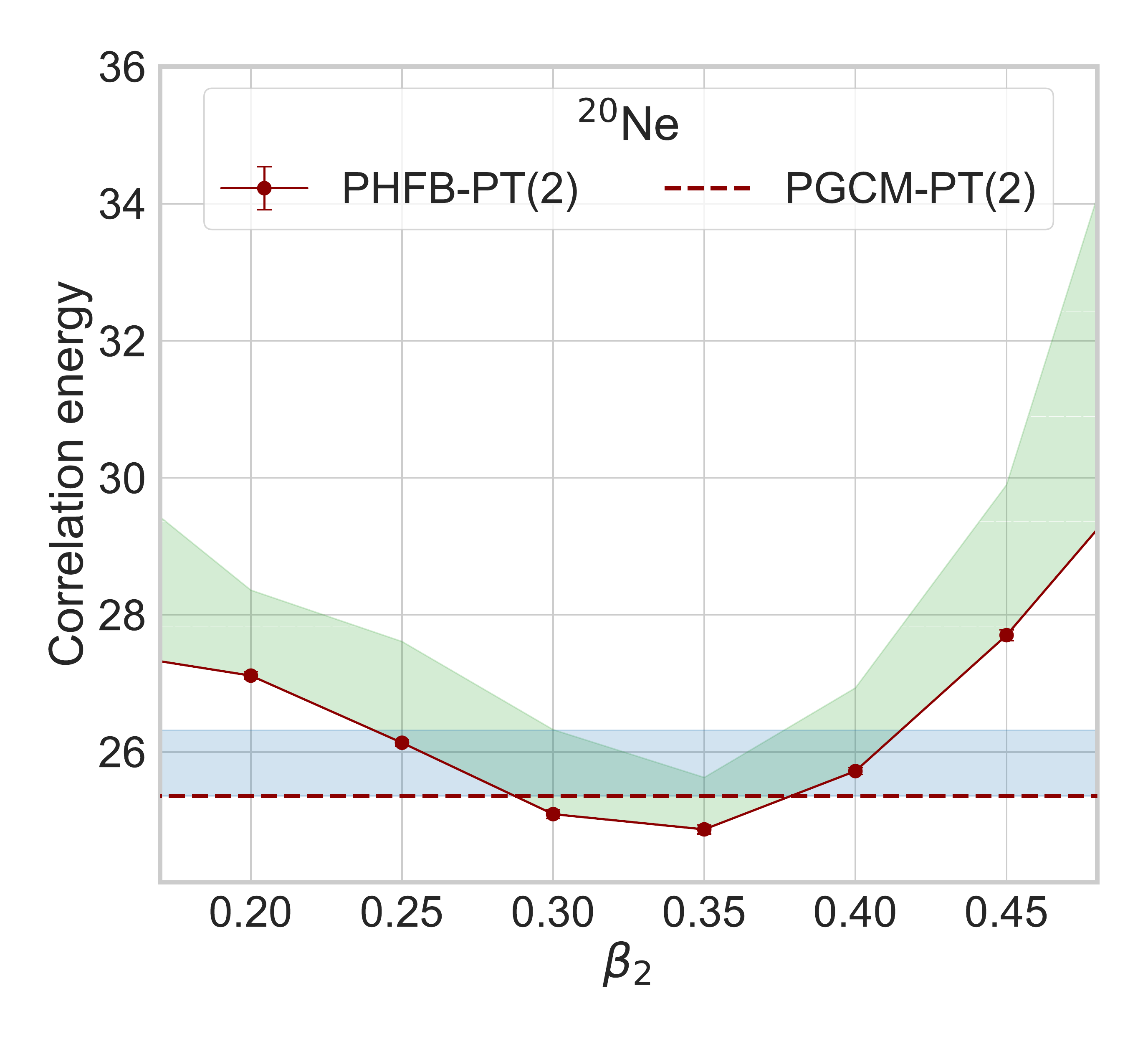}
    \caption{(Color online) Correlation energy in \nucl{Ne}{20} for a complex shift $\gamma=15\,\mathrm{MeV}$. Error bars associated to the effect of the shift correspond to the correlation energy with $\gamma=4\,\mathrm{MeV}$. The calculation is performed with a two-body \(\chi\)EFT Hamiltonian~\cite{Huther_2020,Entem:2017gor}, $\lambda_{\mathrm{srg}}=1.88\,\mathrm{fm}^{-1}$, $\hbar \omega=20\,\mathrm{MeV}$ and $e_{\mathrm{max}}=2$.}
    \label{fig:Ne20_shift_4}
\end{figure}

In practical applications, the optimal shift depends on the interaction, the model space and the system under consideration. As the model space is enlarged, encountering small eigenvalues becomes more probable and the complex shift becomes necessary to smear out the contaminations. A shift $\gamma \in[10,20]$\,MeV is well suited to remove spurious behaviors, with an estimated error of around \(4\%\) on the correlation energy, as can be seen in Fig.~\ref{fig:Ne20_shift_4}. The difference between the results obtained with \(\gamma=15\)\,MeV and \(\gamma=4\)\,MeV\footnote{In the present case, \(\gamma=4\)\,MeV constitutes the lowest value that is empirically found to deliver a controlled numerical result. Below this value, the energy curve displays an erratic behavior due to the occurrence of an intruder state.} is used to estimate the bias due to the shift. Note that PHFB-PT(2) is more sensitive to intruder-state problems than PGCM-PT(2), hence the need to employ a larger shift to smooth out singularities on the energy curve. In practice, it is essential to use the same shift for all quantum states of a given nucleus to obtain a consistent bias in absolute binding energies that will largely cancel out in the excitation spectrum. The development of an extrapolation method towards $\gamma \rightarrow 0$ to correct for the bias due to the complex shift is left for a future study.

\section{Discussion on numerics}
\label{App_numerics}

\subsection{Scaling}
\label{complexity}

Any method to solve $A$-body Schr{\"o}dinger equation's comes with its numerical complexity and memory requirement. For a given basis size \(N\) of the one-body Hilbert space, the naive polynomial scaling of runtime and storage of the methods discussed in the present work is displayed in Tab.~\ref{tab:complexity}. 
These asymptotic values are to be revised when particular symmetries are exploited in the many-body bases (e.g. spherical or axial symmetry reducing the size of the many-body tensors at play).
Moreover, prefactors (ignored here) may play a significant role. Nevertheless, the table gives a fair idea of the asymptotic cost of the different many-body techniques.
\begin{table*}
    \centering
    \begin{tabular}{|l|c|c|c|c|c|c|}
    \cline{1-7}
       \bigstrut Method  & HFB & PGCM & BMBPT(2) & BMBPT(3) & PGCM-PT(2) & FCI \\
     \cline{1-7}
 \bigstrut Runtime & \(O(N^4)\) & \(O(n_{\text{proj}} 
n_\text{gcm}^2 N^4)\)   & \(O(N^5)\) & \(O(N^6)\) 
& \(O(n_{\text{proj}} n_\text{gcm}^2 N^8)\) & \(O(N^{\text{A}})\) \\
       Storage & \(O(N^4)\) & \(O(N^4)\)   & \(O(N^4)\) & \(O(N^4)\) & \(O(n_\text{gcm}^2 N^8)\) & \(O(N^{\text{A}})\) \\
    \cline{1-7}
    \end{tabular}
    \caption{Runtime complexity and storage requirements for various resolution methods of the many-body problem. \(n_{\text{proj}}\) denotes the number of gauge angles used for projections and \(n_\text{gcm}\) the number of states used in the mixing.}
    \label{tab:complexity}
\end{table*}

As an example, the computational cost of each individual matrix element at play in PGCM-PT(2), which requires about 1000 vectorized elementary operations, makes the construction of the matrix the most time-consuming step in the calculation. In other words, the computation of the \(N^8\) matrix elements dictates the overall complexity, given that the (approximate) sparsity of the matrix makes the cost of solving the linear system subleading. Similarly, even if BMBPT(3) has the same storage cost as HFB in principle, symmetry properties of the density matrices are used to drastically reduce the number of matrix elements that are needed at the HFB level. In general, the nominal complexity and storage requirement have to be balanced with the optimizations (vectorization, parallelization, compression techniques) that can be applied for a specific method, and they can play a decisive role in practical applications. Also, shape mixing through PGCM scales quadratically with the number of reference states, i.e. a PGCM (PGCM-PT(2)) calculation with 10 states is 100 times more costly than a PHFB (PHFB-PT(2)) calculation.

\begin{figure}
    \centering
    \includegraphics[width=.5\textwidth]{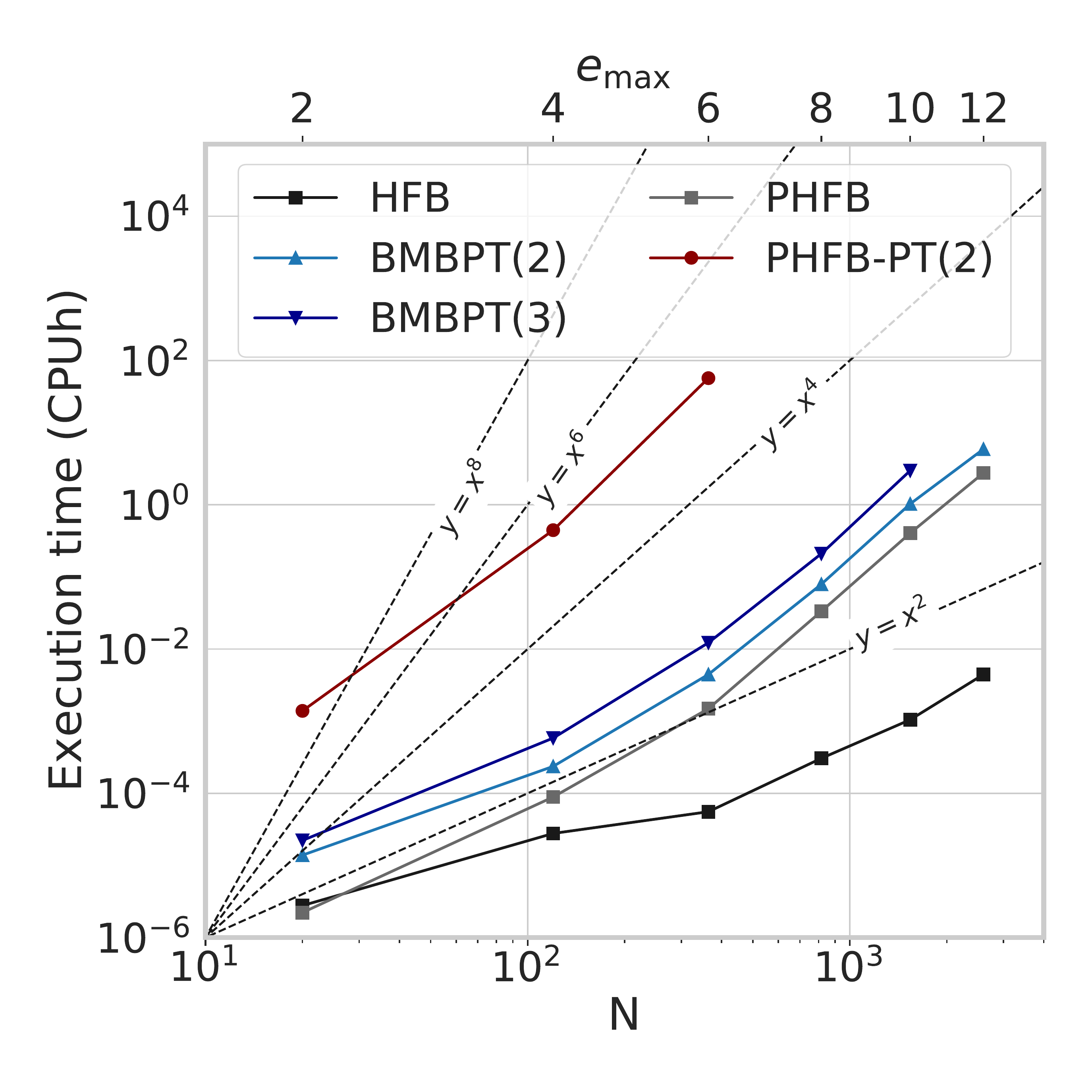}
    \caption{Timing of many-body methods as a function of the basis size of the one body Hilbert space. Projections are performed on total angular momentum \(J\) with 24 gauge angles.}
    \label{fig:complexity}
\end{figure}

A selection of runtimes as a function of the one-body basis dimension is displayed in Fig.~\ref{fig:complexity}. 
For BMBPT(2,3) and PHFB, symmetry properties lower the effective complexity to \(O(N^4)\). The main differences reside in the prefactor, which is, intiutively, larger for BMBPT(3). Note that the normal ordering of the Hamiltonian and the transformation to the quasi-particle basis are included in the runtime estimate.

\subsection{Complexity reduction in PGCM-PT(2)}

The multi-reference PGCM-PT(2) calculation is, in its naive formulation, significantly more costly than its single-reference counterparts. This is mainly due to the redundancies in the visited Hilbert space: many projected quasi-particle configurations play little to no role in the correlation energy or are redundant. This is even more true for large-scale applications where the multi-reference unperturbed state mixes many product HF(B) states. This naturally leads to the idea of reducing the dimensionality of the problem by selecting only relevant configurations (see Refs.~\cite{Tichai_2019,Garniron_2018} for recent applications of this idea in nuclear physics and chemistry). In particular, the application of importance-truncation techniques in the context of non-perturbative methods \cite{porro2021importance} shows promising results that should be applicable to the present problem.

Several procedures to reduce the number of configurations in a controlled way are now briefly introduced, although not all of them have been implemented yet.

\subsubsection{Norm-based importance truncation}
\paragraph{Exact arithmetic}

The norm of a projected configuration is
\begin{equation}
    n^I(p) \equiv \braket{\Omega^I(p)}{\Omega^I(p)} \in [0,1]
\end{equation}
such that a configuration $I$ for which \(n^I(p)=0\) satisfies
\begin{equation}
     \ket{\Omega^I(p)} =0 \, .
\end{equation}
Trivially, a null vector does not contribute to the linear system and can be safely removed from the calculation.

\paragraph{Approximate zeros}

Given a threshold \(\epsilon_n > 0\), the norm-based importance-truncated problem is introduced by removing configurations \(I\) with \(n^I(p) < \epsilon_n\). The exact problem is obtained in the limit \(\epsilon_n=0\). For now, this is the only method that has been implemented and applied to discard configurations in \nucl{O}{18} at \(e_{\text{max}}=6\). Although the number of configurations was divided by two (from $10^6$ to $5 {\cdot} 10^5$) by only keeping configurations whose norm reaches 2\% of the maximal value, the induced error was shown to be less than 1\%. A systematic study of the results obtained via this procedure still remains to be performed.

\subsubsection{Hamiltonian-based importance truncation}

\paragraph{Exact arithmetic}

The Hamiltonian matrix element of a projected configuration reads
\begin{equation}
    h_1^I(p) \equiv \bra{\Omega^I(p)} H_1 \ket{\Theta^{(0)}} \, .
\end{equation}
A configuration $I$ for which \(h_1^I(p)=0\) does not contribute to the linear system nor to the second-order energy, hence it can be safely removed from the calculation.

\paragraph{Approximate zeros}

The Hamiltonian-based importance-truncated problem is introduced by removing configurations \(I\) satisfying \(|h_1^I(p)| < \epsilon_h\), with \(\epsilon_h > 0\). The exact problem is obtained in the limit \(\epsilon_h=0\). 

\subsubsection{Energy-based importance truncation}

The contribution of a configuration \(I\) associated with the vacuum \(\ket{\Phi(q)}\) to the second-order correlation energy is
\begin{equation}
    e^{(2)I}(q) = h_1^{I*}(q) a^{I}(q) \, .
\end{equation}
A configuration \((q,I)\) for which \(e^{(2)I}(q)=0\) does not contribute to the correlation energy\footnote{Such a configuration might still contribute indirectly by influencing the value of the other coefficients \(\{a^J(q')\}\).}. Removing configurations based on the size of their contribution to the correlation energy corresponds to the method advocated in Refs.~\cite{Tichai_2019,porro2021importance}. The method is expected to lead to a substantial gain for a negligible error on the energy, although the impact on other observables must be checked as well. Of course, computing $e^{(2)I}(q)$ requires to solve the problem in the first place and is thus impractical. The idea is thus to evaluate the importance of a given configuration $(q,I)$ by calculating an approximation to $e^{(2)I}(q)$ at a significantly reduced cost, which can typically be achieved by using BMBPT(2) based on the HFB vacuum $| \Phi(q) \rangle$.  



\end{appendices}

\bibliography{bibliography.bib}

\end{document}